\documentclass[11pt]{article}
\pdfoutput=1
\usepackage[utf8]{inputenc}
\usepackage{multirow, amsmath, amsfonts, amssymb}
\usepackage{comment}
\usepackage{graphicx}
\usepackage{psfrag}
\usepackage{amsthm}
\usepackage[dvipsnames,svgnames,table]{xcolor}
\usepackage{enumerate,arydshln,soul,slashed,subfig,tabularx,caption}
\usepackage{mathrsfs,a4wide,tikz,tikz-cd}
\usetikzlibrary{shapes.geometric}
\usepackage{tcolorbox,listings,jheppub,hyperref}
\usepackage[nameinlink, english]{cleveref}
\usepackage{setspace}
\usepackage{color}
\definecolor{codebg}{rgb}{0.97,0.97,0.97}
\definecolor{codeframe}{rgb}{0.85,0.85,0.85}
\definecolor{keyword}{rgb}{0.15,0.15,0.7}
\definecolor{string}{rgb}{0.65,0.15,0.15}
\definecolor{comment}{rgb}{0.25,0.5,0.35}
\definecolor{number}{rgb}{0.55,0.0,0.55}
\definecolor{identifier}{rgb}{0.1,0.1,0.1}
\definecolor{dark-red}{rgb}{0.7,0,0}
\definecolor{dark-green}{rgb}{0.1,0.4,0}
\definecolor{dark-blue}{rgb}{0.3,0.3,0.7}

\hypersetup{colorlinks=true,
	linkcolor=dark-blue,
	citecolor=dark-red,
	urlcolor=dark-green,
	linktoc=page}

\newcommand\SmallMatrix[1]{{%
\text{\tiny$\arraycolsep=1\arraycolsep\ensuremath{\begin{pmatrix}#1\end{pmatrix}}$}}}
\newcommand{\C}{\mathbb{C}}
\newcommand{\R}{\mathbb{R}}
\newcommand{\Q}{\mathbb{Q}}
\newcommand{\Z}{\mathbb{Z}}
\newcommand{\Spin}{\mathrm{Spin}}

\numberwithin{equation}{section}

\theoremstyle{remark}

\crefname{appendix}{Appendix}{Appendices}
\crefname{section}{Section}{Sections}

\title{\centering Non-abelian asymmetric orbifolds with vanishing one-loop vacuum energy}
\author[1]{Bernardo Fraiman,}
\author[2,3]{Vittorio Larotonda}
\author[1]{and Michelangelo Tartaglia}
\affiliation[1]{Instituto de Física Teórica IFT-UAM/CSIC, C/ Nicolas Cabrera 13-15, Campus de Cantoblanco, 28049 Madrid, Spain}
\affiliation[2]{Dipartimento di Fisica e Astronomia, Universit\`{a} di Bologna, via Irnerio 46, Bologna, Italy}
\affiliation[3]{INFN, Sezione di Bologna, viale Berti Pichat 6/2, Bologna, Italy}
\emailAdd{bernardo.fraiman@csic.es}
\emailAdd{vittorio.larotonda@unibo.it}
\emailAdd{michelangelo.tartaglia@estudiante.uam.es}
\abstract{We present a partial classification of four-dimensional, non-supersymmetric Type II toroidal orbifolds with non-abelian point groups of rotations, and vanishing vacuum energy at one loop in string perturbation theory. The classification is complete within a class of such orbifolds with the restriction that the point group only acts on a $T^5$ inside the full internal $T^6$. By studying their decompactification limits along the leftover $S^1$, we see that the 17 solutions we find can alternatively be obtained as Scherk-Schwarz compactifications of some parent asymmetric orbifold. Along the way, to ensure the absence of anomalies, we compute bordism groups $\Omega_3^\Spin(BG)$ for a variety of non-abelian crystallographic groups $G$.}
\hypersetup{pageanchor=false}
\makeatletter
\let\old@fpheader\@fpheader
\preprint{ {\tt IFT-UAM/CSIC-26-74}\\}
\hypersetup{
    pdftitle={Non-abelian asymmetric orbifolds with vanishing one-loop vacuum energy},
    pdfauthor={Bernardo Fraiman, Vittorio Larotonda, Michelangelo Tartaglia},
    pdfsubject={asymmetric orbifolds}}

\begin{document}

\maketitle

\section{Introduction and summary of results}

String theoretic models with a mechanism that suppresses the cosmological constant are of clear phenomenological interest, given the measured smallness of the cosmological constant in our universe.

Spacetime supersymmetry is a straightforward way to achieve this: non-renormalization theorems ensure that a supersymmetric Minkowski vacuum has vanishing cosmological constant at all orders in perturbation theory \cite{Martinec:1986wa,Witten:2012bh,Witten:2013cia}.
In general, in string perturbation theory, the vacuum energy has an expansion in the string coupling $g_s$:
\begin{equation}\label{eq:vacuum_energy_expansion}
       V= \sum_{g=0}^{\infty} V_g\, g_s^{2g-2}\,.
\end{equation}
While the zeroth-order term vanishes identically due to $SL(2,\C)$ invariance, in the absence of supersymmetry there is in general no mechanism to protect any other coefficient $V_g$.

In this paper we explore models in which at least the $g=1$ term
\begin{equation}\label{eq:vacuum_energy_one_loop}
    V_1 = \int_\mathcal{F} \frac{d^2\tau}{(\text{Im}{\tau})^2} \mathcal{Z}(\tau,\bar{\tau})\,,
\end{equation}
vanishes. Here $\mathcal{Z}(\tau,\bar{\tau})$ is the one-loop partition function of the worldsheet CFT, and $\mathcal{F}$ is the fundamental domain of $SL(2;\Z)$:
\begin{equation}
    \mathcal{F} = \left\{ \tau \in \C \,|\,\tau_1 \in [-\tfrac{1}{2},\tfrac{1}{2}], \tau_2>0,|\tau|>1  \right\}\,.
\end{equation}
There are two ways in which one can have \eqref{eq:vacuum_energy_one_loop} vanish:
\begin{itemize}
    \item The integrand $\mathcal{Z}(\tau,\bar{\tau})$ may be non-zero, but the full integral over the moduli space in \eqref{eq:vacuum_energy_one_loop} still vanishes. This is Atkin-Lehner symmetry \cite{Moore:1987ue}, which however has some limitations: it has been achieved only in 2D target spaces \cite{Dienes:1990qh}, and it has proven to be impossible to realize for toroidal orbifolds in $D>2$ \cite{Gannon:1992su}.
    \item The function $\mathcal{Z}(\tau,\bar{\tau})$ vanishes pointwise for all $\tau$. This happens in supersymmetric theories, and it is the property we want to engineer in this work in the absence of SUSY.
\end{itemize}

The mechanism we use to obtain a vanishing $V_1$ was pioneered in \cite{Kachru:1998hd}, and further explored in \cite{Kachru:1998yy, Shiu:1998he, Blumenhagen:1998uf, Angelantonj:1999gm, Angelantonj:2004cm, Satoh:2015nlc, Aoyama:2020aaw,Larotonda:2026hxy}. Many other approaches to obtain similar suppression have also been studied, see \cite{Angelantonj:2003hr,Abel:2017vos, Satoh:2021nfu,Dudas:2025yqm}.
The main idea in \cite{Kachru:1998hd} is the following: a worldsheet CFT,  obtained as an orbifold of some other ``parent'' CFT by a finite group $G$, has a partition function that can be written as a sum 
\begin{equation}\label{eq:orbifold_partition_function}
    \mathcal{Z}_{\text{orbifold}}(\tau,\bar{\tau}) = \frac{1}{N} \sum_{(f,g)} \mathcal{Z}[f,g]\quad \text{with} \quad  f,g\in G\quad\text{and}\quad fgf^{-1}g^{-1}=1\,,
\end{equation}
where $\mathcal{Z}[f,g]$ denotes the torus partition function of the parent CFT with holonomies of $f,g\in G$ along the two independent cycles of the torus, and $N$ is a normalization constant.

The sum is only over commuting pairs since what \eqref{eq:orbifold_partition_function} represents is a gauging by $G$: we are summing over all $G$-bundles over the worldsheet torus, which for a discrete group $G$ are classified by maps in $\text{Hom}(\pi_1(T^2),G)\simeq \text{Hom}(\Z^2,G)$.

Each summand in \eqref{eq:orbifold_partition_function} has a Hilbert space interpretation as the trace of action of the element $g$ in the $f$-twisted Hilbert space
\begin{equation}
    \mathcal{Z}[f,g]=\text{Tr}_{\mathcal{H}_f} (q^{L_0-c_L/24} \bar{q}^{\bar{L}_0-c_R/24}\hat{g}\, P_{GSO})\,.
\end{equation}
If there is a spacetime supercharge operator $Q$ in the $f$-twisted sector which commutes with $\hat{g}$, the partition $\mathcal{Z}[f,g]$ will vanish because of Bose-Fermi cancellation \cite{GrootNibbelink:2017luf}.

Additionally, this guarantees that the orbifold theory is tachyon-free, since a tachyon in the spectrum would lead to a divergence in the one-loop vacuum energy \cite{Larotonda:2026hxy}.

The insight in \cite{Kachru:1998hd} is that one can construct an orbifold CFT such that each commuting pair $(f,g)$ preserves a supercharge, but crucially \textit{not the same in all sectors}. Then spacetime SUSY is broken explicitly, since no supercharge survives in the 4D spectrum, but each summand is ``like that of a supersymmetric theory", and thus vanishes.

In practice, to find such models, one notices that a subset of commuting pairs is certainly given by
\begin{equation}\label{eq:trivialpair}
    (1,g) \quad \text{for any  } \, g \in G\,,
\end{equation}
so we need to impose that every element $g \in G$ preserves some supercharge, without any supercharge being preserved by the whole group.
\\This is the core of the representation theory problem we solve in \hyperref[sec:rep_th]{Section 3}.

If this condition is met, then all of the partition function sectors
\begin{equation}
    \mathcal{Z}[g^a,g^b] =0 \,
\end{equation}
by modular covariance, since $(g^a,g^b)$ is in the modular orbit of $(1,g)$.

If there are no other commuting pairs, then the model has vanishing vacuum energy at one loop. This was the original claim for the model in \cite{Kachru:1998hd}, which however is not quite true, as pointed out already in \cite{Harvey:1998rc}: the original orbifold is a $T^6/(\Z_2\times \Z_2)$ model, and the two generators clearly commute. The claim that they do not commute is based on an invalid equivalence to an orbifold of a non-compact $\R^6$ CFT by an extension of $\Z_2 \times \Z_2$ by the discrete translation describing the torus lattice. This always holds for symmetric orbifolds, where the action one quotients by is a symmetry for every value of the radii. Asymmetric orbifold actions are generally symmetries only at special loci in Narain moduli space. The simplest example of this phenomenon is T-duality, which in our language is a $\Z_2$ asymmetric action $X_R^\mu \mapsto -X_R^\mu$, and is a symmetry of the theory only at the self-dual radius $R = \sqrt{\alpha'}$.

The point is that for asymmetric orbifolds one has to directly work at the level of the $T^6/G$ orbifold, and consider in \eqref{eq:orbifold_partition_function} all pairs that commute at the space group level, i.e.~including translations.

Most of the non-abelian groups we end up needing to consider have relatively few additional commuting pairs other than those of the form \eqref{eq:trivialpair}, but they do have to be checked separately. As a rule of thumb, the higher the order of the group, the more this additional check becomes restrictive.

A full classification of such toroidal orbifolds that have an abelian point group (i.e.~the group of rotations) was proposed in \cite{Larotonda:2026hxy}. In this work we aim to extend this classification to a large class of orbifolds which are non-abelian already at the level of the point group. We approach this challenge with a simplifying assumption: we consider only a subclass of $T^6$ orbifold compactifications acting on an $S^1$ as a pure geometric shift, while having a non-trivial $SO(5)\subset SO(6)$ point group action on a $T^5$. Focusing on these actions simplifies the computational effort considerably, since we are essentially only looking at automorphism groups of rank-5 lattices, which are far fewer than their rank 6 counterparts. Apart from this, there is also a useful conceptual simplification: being able to use geometric shifts over the ``base" $S^1$ allows us to choose an arbitrarily large radius for the last coordinate, which avoids the accidental restoration of supersymmetry via massless gravitini in the twisted sectors, see e.g.~\cite{Florakis:2017zep,Baykara:2024vss}.

While we do not expect any significant conceptual complication arising, we leave the classification of genuine $SO(6)$ orbifold actions over $T^6$ for future work.

Furthermore, we are mostly concerned with the classification of point groups $P_G$ of rotations, since they are those that act on the 4-dimensional supercharges. We therefore do not distinguish between the a priori very large number of models arising from the same point group equipped with different choices of shift. We adopt a minimalist approach and select the simplest choice of shift that ensures our constructions yield non-supersymmetric vacua with vanishing one-loop vacuum energy. In most cases, it will be enough to use the geometric shift on the $S^1$ we purposefully left invariant. We will comment explicitly later where this is not enough.

The classification effort for non-abelian groups requires considerably more involved group theoretic tools. Representation theory for any abelian group is simple enough that it was possible in \cite{Larotonda:2026hxy} to limit the allowed abelian point groups for any orbifold CFT, not necessarily toroidal, to a small finite list. In the same reference, it was also shown that there is no bound one can in general put on the order of the allowed non-abelian groups on purely group theoretic grounds. This was done by explicitly finding a solution for $G=D_{2n}$, a dihedral group of arbitrarily high order $2n$. To obtain a finite list, we restrict ourselves to the context of toroidal orbifolds, where the classification of crystallographic symmetries puts strong restrictions on the list of groups we are even allowed to consider in the first place.

\begin{table}[t] \label{tab:results}
\renewcommand{\arraystretch}{1.3}\centering
\begin{tabular}{|
@{}>{\,}c<{\,}@{}|
@{}>{\,$}c<{$\,}@{}|
@{}>{\,$}c<{$\,}@{}|
@{}>{\,$}c<{$\,}@{}|
>{$}c<{$}|} \hline
 \text{Id(G)} & {G} & G_F & G_B &  \text{\#} \\ \hline
 {$[6,1]$} & \hyperref[sec:S3]{S_3} & { \Z_2}\oplus {S_3}& 1 \oplus { S_3 }  & 1  \\ \hline
 {$[8,4]$} &  \hyperref[sec:Q8]{Q_8} & {(\Z_2 \times \Z_2)}\oplus Q_8 & {\Z_2}\oplus Q_8 & 3 \\\hline
 \multirow{2}{*}{{$[12,1]$}} &  \text{\multirow{2}{*}{\hyperref[sec:Dic12]{$Dic_{12}$}}} &\Z_4\oplus  S_3  & \Z_4 \oplus S_3 & 1  \\
 && Dic_{12}\oplus { S_3 } & Dic_{12}\oplus { S_3 } & 1 \\\hline
 {$[12,4]$} &  \hyperref[sec:D12]{D_{12}} & {(\Z_2 \times \Z_2)}\oplus  { S_3 }&{\Z_2}\oplus  S_3 & 1 \\\hline
 \multirow{2}{*}{{$[18,3]$}} & \text{\multirow{2}{*}{\hyperref[sec:S3Z3]{$S_3 \times  \Z_3$}}} & 
 \text{\multirow{2}{*}{${ \Z_6 }\oplus { S_3 }$}}& {\Z_3}\oplus{ S_3 }&1 \\
 &&& { \Z_6}\oplus{ S_3 }&1 \\\hline
 \multirow{2}{*}{{$[24,4]$}} & \text{\multirow{2}{*}{\hyperref[sec:Dic24]{$Dic_{24}$}}}& Q_8\oplus  S_3& Q_8\oplus S_3 &1\\
 &&  Q_8\oplus  (S_3 \times \Z_2)& { Q_8 } \oplus S_3&1 \\\hline
 {$[24,5]$} & \hyperref[sec:S3Z4]{S_3 \times  \Z_4 } &(\Z_4 \times \Z_2) \oplus  S_3 &  {\Z_4}\oplus S_3 &1 \\\hline
 {$[36,7]$} & \hyperref[sec:G736]{G^7_{36}}
 & Dic_{12}\oplus S_3 & Dic_{12}\oplus { S_3 }& 1  \\\hline
{$[36,12]$} & \hyperref[sec:S3Z6]{ S_3 \times  \Z_6 } & { (\Z_6 \times \Z_2)} \oplus{ S_3 }& { \Z_6}\oplus { S_3 }& 1 \\\hline
{$[48,40]$} &  \hyperref[sec:S3Q8]{ S_3 \times  Q_8 } & (Q_8 \times \Z_2)\oplus { S_3 }& Q_8\oplus { S_3 }& 1  \\\hline
 {$[72,20]$} & \hyperref[sec:S3Dic12]{ S_3 \times  Dic_{12} } & (Dic_{12} \times \Z_2)\oplus{ S_3 } & Dic_{12}\oplus { S_3 }& 1 \\\hline
 {$[144,128]$} & \hyperref[sec:S3SL23]{ S_3  \times SL(2,3)} & (SL(2,3) \times \Z_2)\oplus { S_3 }& {SL(2,3)}\oplus { S_3 }& 1  \\\hline
\end{tabular}
\caption{A summary of our solutions. The full groups $G$ are listed, together with their ID in GAP and their decomposition into left and right movers, both at the level of fermions and bosons. The last column indicates the number of inequivalent orbifolds for each given $G$.
The three $Q_8$ orbifolds have a non-trivial action on a $T^4$, leaving a $T^2$ invariant, while all the others act on a $T^5$, leaving only an $S^1$ invariant. }
\label{tab:listFinal}
\end{table}

In particular, symmetries of lattices of rank 5 have been fully classified \cite{Plesken2000CountingCG}, and the full classification can be accessed through the software CARAT \cite{Plesken:zm0025}.
Let us establish our conventions for finite groups we use in this work. 
\begin{itemize}
    \item $S_n$ is the symmetric group over $n$ objects.
    \item $Q_8$ is the quaternion group.
    \item $D_{2n}$ is the dihedral group of order $2n$.\footnote{Some authors use the notation $D_n$ to emphasize that it is the symmetry group of a regular $n$-gon. Throughout the paper we stick to the convention of using subscripts for the order of the group, except for $S_n$.}
    \item $Dic_{4n}$ stands for the dicyclic group of order $4n$.
    \item $SL(2,3)= SL(2;\Z_3)$ is the special linear group of rank two over $\Z_3$.
    
\end{itemize}
Presentations of these groups are provided in \hyperref[sec:anomalies]{Section 4} together with their representation theory, which we will use to compute the anomalies of our orbifolds.

Similar tools to ours were employed in \cite{GrootNibbelink:2017luf} to exclude the existence of heterotic toroidal orbifolds with vanishing one-loop vacuum energy, and as remarked in \cite{Larotonda:2026hxy}, this result also excludes symmetric Type II orbifolds. This is because imposing the action to be left/right symmetric is entirely equivalent, from the group theory point of view, to just having an action on the e.g.~left-moving supercharges.

We are therefore forced to use asymmetric Type II orbifolds, whose main features and technical details are reviewed in \hyperref[sec:as_orbifolds]{Section 2}.

For now, this introduces two important differences. First, the action on left- and right-moving supercharges can be defined independently from one another, giving considerably more freedom. Second, asymmetric actions can in general give rise to massless gravitini in the twisted sectors: this would ``accidentally" restore SUSY, and it is something we want to avoid, by adding suitable shifts to the orbifold action on bosons when needed. 

The strategy we apply to find these orbifolds is schematically as follows:

\begin{itemize}
    \item[1.]We define a set of conditions that the action of the orbifold group $G$ on supercharges needs to satisfy for the partition function to vanish and SUSY to be explicitly broken at least in the untwisted sector.
    \item[2.]Using the data from CARAT we build a list of the possible candidate $G$. This is the exhaustive list of all possible groups that can be the spin lift of a crystallographic group acting asymmetrically on a 6D lattice.
    \item[3.]For all these candidate groups $G$ we find all fermionic actions $\rho_F(G)$ satisfying the properties defined in point 1.
    \item[4.]For all these representations, we compute the induced action $\rho_B(G)$ on bosons and make sure that it is crystallographic using CARAT again.
    \item[5.]If a representation passes all the above checks, we impose further consistency conditions: anomaly cancellation and the absence of massless twisted gravitini.
    
\end{itemize}

The results of this search can be found in \hyperref[sec:5Dresults]{Section 5} and are summarized in \hyperref[tab:results]{Table 1}.

One can see that our conditions are quite restrictive, and only lead to a handful of allowed point groups. All of the orbifolds in \hyperref[tab:results]{Table 1} enjoy one additional property: we were able to add shifts in such a way that the full orbifold group $G$ is isomorphic to the point group $P_G$. This ensures that our conditions are enough without further checks.

We also have a large class of candidates, which we discuss further in \hyperref[sec:5Dresults]{Section 5}, for which it is in principle possible to add shifts so that the space group gets enhanced and the orbifold sum \eqref{eq:orbifold_partition_function} is rearranged, both removing and adding sectors, in such a way that the vacuum energy still vanishes. This is precisely what happens in the models of \cite{Larotonda:2026hxy}, for which the abelian nature of the point group made it necessary to add such shifts. We expect this not to be possible in general for all models in this list. Since it introduces considerable complications both computationally and from the point of view of global anomalies, we do not attempt to add this kind of shift.

The rest of the paper is organized as follows. In \hyperref[sec:as_orbifolds]{Section 2} we review the main techniques associated with asymmetric orbifolds, focusing on the extra subtleties coming with their non-abelian version. In \hyperref[sec:rep_th]{Section 3} we turn the search for non-supersymmetric orbifolds with vanishing vacuum energy into a precise representation theoretic problem, and we describe the strategy sketched above, that we use to fully solve the problem. In \hyperref[sec:anomalies]{Section 4} we compute the global anomalies associated with non-abelian orbifolds through the modern perspective of cobordism theory. Finally, in \hyperref[sec:5Dresults]{Section 5} we present some of the new models in great detail. In the Appendices we develop in more detail some of the mathematical tools needed for our analysis. 

\section{Asymmetric orbifolds}\label{sec:as_orbifolds}

In this Section, we review the essential tools for the study of toroidal orbifolds in Type II string theory. For more details, see \cite{Larotonda:2026hxy} and references therein.

A Type II compactification on $T^d$ is fully specified by a choice of Narain lattice, namely an embedding $\Gamma^{d,d} \hookrightarrow \R^{d,d}$ of an even, self-dual lattice of signature $(d,d)$.

Points in this lattice represent the quantized internal momenta $(p_L,p_R)$, defined in terms of Kaluza-Klein momentum and winding of the internal coordinates. The inner product is defined through the indefinite norm $p_R^2-p_L^2$. The orbifold theory is obtained by taking the quotient with respect to a subgroup $G$ of the automorphisms of the Narain lattice \cite{NARAIN198641}. In particular, we focus on crystalline symmetries, for which each element $g \in G$ can be factorized as
\begin{equation}\label{eq:Gaction}
    g | p_L,p_R \rangle = e^{2 \pi i \left(-v_L \cdot p_L + v_R \cdot p_R \right)} | \theta_L \, p_L, \theta_R \, p_R \rangle\,,
\end{equation}
where $\theta_L$ and $\theta_R$ are both lattice automorphisms of $\Gamma^{d,d}$ usually called \textit{twists}. $\theta_L$ acts only on the left-moving sector, while $\theta_R$ acts on the right movers. The additional phase is encoded by a $(d+d)$-dimensional vector $v = (v_L,v_R)$ known as the \textit{shift vector}. In the context of toroidal orbifolds, the full group generated by twists and shifts is often called the \textit{space group}, while the group generated only by the twists is dubbed the \textit{point group}. 
When $\theta_L = \theta_R$ and $v_L = v_R$ for every element in the group $G$, the action is called symmetric, and the resulting model still has an interpretation as a sigma model of strings propagating on the singular space $T^d/G$. This is because if the action on the left and right-moving bosons $X_L^\mu$ and $X_R^\mu$ is the same, it descends naturally to a well-defined action on the spacetime coordinates $X^\mu = X_L^\mu + X_R^\mu$.

If $\theta_L \neq \theta_R$ for any element, we lose the geometric interpretation, but the model nonetheless makes sense as a quotient of the worldsheet CFTs. We denote the left-moving bosonic point group $G_B^L$ as the abstract group generated by the twists $\theta_L$, and similarly $G_B^R$ for right movers.

Then by construction both $G^L_B$ and $G^R_B$ are subgroups of $GL(d;\Z)$. We will consider only group actions for which the point group is finite. This allows us to use results from the classification of symmetries of crystals, which are in one-to-one correspondence with subgroups of the relatively short list of maximal finite subgroups of $GL(d;\Z).$

Any given element $\theta_{L/R}$ is represented by an integral, and hence real, matrix. Then its eigenvalues come in complex conjugate pairs of phases $e^{2 \pi i \theta_i^{L/R}}$, and we can represent the element as a \textit{twist vector}
\begin{equation}
\left\{
\begin{aligned}
    \Vec{\theta}^{L/R} &= (\theta^{L/R}_1,...,\theta^{L/R}_{d/2})\, \qquad &\text{for $d$ even}\,,\\
    \Vec{\theta}^{L/R} &= (\theta^{L/R}_1,...,\theta^{L/R}_{(d-1)/2},0)\, \qquad &\text{for $d$ odd}\,.
\end{aligned}
\right.
\end{equation}

There are consistency conditions on the choices of twist $\theta$ and shift vector $v$, first among them modular invariance.

For abelian orbifolds, modular invariance is equivalent to level matching \cite{Narain:1986qm, Freed:1987qk}, which requires that in every sector twisted by an element $g$ of order $N$ the energy levels obey
\begin{equation} \label{eq:levelMatch}
    N (E_L - E_R) \in \Z\,,
\end{equation}
for any pair of left and right-moving states. We are going to consider non-abelian orbifolds in this work, but a subset of the conditions for modular invariance of a $T^d/G$ are given by imposing it on all abelian subgroups $A \subset G$. In particular, on the cyclic subgroups generated by each element. In Type II, \eqref{eq:levelMatch} translates to a condition on the twist and shift vector \cite{Baykara:2024vss}
\begin{equation}\label{eq:modInv}
    \sum_i \{ \theta^R_i \} - \sum_i \{ \theta^L_i \}  + (v^*_R)^2-(v^*_L)^2 \, \in \, \frac{2\Z}{N} \,.
\end{equation}
We will impose this on each element of the non-abelian group $G$. When this is successfully done, we say that the ``abelian anomaly" is cancelled. In contrast, we will refer as the ``non-abelian anomaly" to any anomaly that may still be present after we ensured level matching.

In fact, it is known that for non-abelian groups there may be additional global anomalies \cite{Freed:1987qk} not induced by any abelian subgroup. From the worldsheet point of view these arise because in non-abelian orbifolds, states do not quite arrange themselves into sectors twisted by an element $g \in G$, but rather in a whole equivalence class under conjugation, $[g] = \{cgc^{-1} | c \in G\}$.

This can be seen, for example, in the following way: consider a state $\psi$ which is invariant under the action of some element $g$. We can act on $\psi$ with some group element $c$ which does not commute with $g$. What one obtains is not a $g$-invariant state anymore, but rather a state $c\psi$ which is invariant under the conjugate element $cgc^{-1}$.

Concretely, this means that there should be no phase difference in the partition function \eqref{eq:orbifold_partition_function} when considering insertions which are obtained by conjugating with a common element, i.e.
\begin{equation}
    \mathcal{Z}[f,g] = \mathcal{Z}[cfc^{-1},cgc^{-1}]\,.
\end{equation}
This phase assignment may not be done in a way that is consistent with modular invariance. In fact one can have triplets $(f,g,c)$ for which a modular transformation sends
\begin{equation} \label{eq:nonAbmodular}
    (f,g) \xrightarrow[]{SL(2;\Z)}(f^ag^b,f^cg^d) = (cfc^{-1},cgc^{-1})\,.
\end{equation}
This modular transformation generically gives the partition function a phase, which we then need to impose to be trivial. The phase can be calculated directly \cite{Freed:1987qk} when the explicit representation in which all chiral fields transform is known.

One can however express this kind of anomaly, together with more general ones, in modern cobordism language. We do this in \hyperref[sec:anomalies]{Section 4} to ensure the absence of all possible anomalies in modular invariance.

We are always going to work at points in the Narain moduli space which have enhanced discrete symmetry.
Given a rank $d$ even lattice $\Lambda$, one can build a Narain lattice $\Gamma^{d,d}$ through the gluing procedure we review in \hyperref[app:cyclotomic]{Appendix A}:

\begin{equation}\label{eq:extraSymm}
    \Gamma^{d,d}(\Lambda) = \left\{ (p_L,p_R) \in \Lambda^* \times \Lambda^* \, \bigg{|} \, p_L-p_R \in \Lambda   \right\} \,,
\end{equation}
where $\Lambda^*$ denotes the dual lattice to $\Lambda$. A typical case is obtained by taking $\Lambda = \Lambda_R(\mathfrak{g})$, the root lattice of a simply laced Lie algebra $\mathfrak{g}$ of ADE type \cite{Ginsparg:1986bx,Dabholkar:1998kv}.
Then by construction, the Weyl group $W(\mathfrak{g)}$ is a symmetry. We will see in \hyperref[sec:5Dresults]{Section 5} that most lattices with enhanced symmetry are of this form.
The torus lattice defining $T^6 = \R^6/\Lambda$ is obtained as the pure winding component
\begin{equation}\label{eq:physlattice}
    \Lambda = \left\{2 \pi \sqrt{\frac{\alpha'}{2}} (p_L-p_R) \, \bigg{|} \, (p_L,p_R) \in \Gamma^{d,d} \right\}\,.
\end{equation}
In the examples we just discussed, this is the root lattice of $\mathfrak{g}$, up to overall scaling.

There is one more subtlety left to discuss when considering non-supersymmetric models. We choose representations of $G$ that break all SUSY in the untwisted sector, by explicitly projecting out the gravitini from the spectrum. In asymmetric orbifolds, there is still the possibility that massless gravitini may emerge from twisted sectors, leading to an accidental restoration of SUSY. As shown in \cite{Larotonda:2026hxy}, there are strict restrictions on when this accidental enhancement can happen.

In light-cone gauge language, a 4D gravitino can only arise as a tensor product of a massless $SO(2)$ vector, say on the left, and a massless $SO(2)$ spinor, say on the right. For ease of notation, for the rest of this Section we will assume the vector comes from the left movers. Obviously the discussion is totally symmetric under $L\leftrightarrow R$.  For a group element $g = (\theta,v)$ the mass of a left-moving spacetime vector in the $g$-twisted sector is given as:
\begin{equation}\label{eq:Hvec}
    H_L^\text{vect} = \frac{(p_L + v_L)^2}{2} + \sum_i |\theta_{L,i}|.
\end{equation}
where $p_L$ is the momentum in the internal $T^6$ components. Internal momenta live in the dual of the invariant lattice associated with the twist $\theta$, see e.g.~\cite{Dabholkar:1998kv}:
\begin{equation}
    (p_L,p_R) \in I^*_\theta \qquad \text{for }\quad I_\theta = \bigg{\{} \, (p_L,p_R) \, \bigg{|} \, \theta | p_L,p_R \rangle = |p_L,p_R \rangle \, \bigg{\}}\,,
\end{equation}
so one can see that, to have a massless vector, $\theta_L$ must be trivial and that $(v_L,v_R) \in I_\theta^*$, so that there is some internal momentum state that can kill the first term in \eqref{eq:Hvec}.
\\The mass of a right-moving $SO(2)$ spinor is given as
\begin{equation}\label{eq:Hspinor}
    H_R^\text{spinor} = \frac{(p_R + v_R)^2}{2} + \sum_i\frac{1}{2}(1 + \epsilon_i)|\theta_i|\,,
\end{equation}
for some choices of sign $\epsilon_i = \pm 1$. This imposes no additional restriction on the twist $\theta_R$, since there is always a choice of sign that sets the second term in \eqref{eq:Hspinor} to zero.

If there is an element $g\in G$ such that $\theta_L=1$ we have two avenues to get rid of the corresponding gravitino. First, to construct the orbifold spectrum, as can be seen by the sum \eqref{eq:orbifold_partition_function} one has to project on $G$-invariant subspaces also in the twisted sectors. More precisely, because of the condition of commuting pairs, in the $g$-twisted sector one has to project on elements left invariant by the centralizer of $g$, $\mathcal{C}(g)\subset G$:
\begin{equation}
    \begin{aligned}
        \mathcal{Z}_{g\text{-twisted}} = \sum_{[h,g]=1}\mathcal{Z}[g,h] = \sum_{h\in \mathcal{C}(g)} \text{Tr}_{\mathcal{H}_g}(\hat{h} \,q^{L_0-c_L/24} \bar{q}^{\bar{L}_0-c_R/24}\, P_{GSO}) =\\
        = |\mathcal{C}(g)|\text{Tr}_{\mathcal{H}_g}(P_{C(g)} \,q^{L_0-c_L/24} \bar{q}^{\bar{L}_0-c_R/24}\, P_{GSO})\,,
    \end{aligned}
\end{equation}
where $P_H$ denotes the projector on the space invariant under a subgroup $H$. This may not be enough, since for some groups it can happen that the centralizer of some element $g$ is generated by $g$ itself: $\mathcal{C}(g) = \langle g \rangle$. If this is the case, since the twist $\theta_L$ needs to be trivial to potentially have a gravitino in the first place, then the projection is also trivial on the left and the gravitino survives in the spectrum. This is for example what trivially happens when the orbifold group is cyclic, see e.g.~\cite{Florakis:2017zep,Baykara:2024vss} for models with this enhancement. On the other hand, if there is any element in $\mathcal{C}(g)$ such that its left-moving twist is not trivial, it will project out the $SO(2)$ vector out of the spectrum, and any gravitino accordingly.

As a simple example, in our $G=Q_8$ models below the potential gravitini would appear in the sector twisted by the element $-1\in Q_8$, since it acts trivially on left moving bosons. Since $-1$ is central in $Q_8$, the centralizer is the whole group: $\mathcal{C}(-1)= Q_8$, and the gravitino is clearly projected out since some element acts non-trivially either on the left or on the right.

If this projection does not suffice, we need to add shifts to our orbifold action, to lift either of the masses \eqref{eq:Hvec}, \eqref{eq:Hspinor}.

To summarize, in order not to have massless gravitini in the spectrum we impose the additional condition: for any element $g=(\theta,v)$ for which say $\theta_L$ is trivial, either $\mathcal{C}(g)$ contains an element with a non-trivial left-moving twist, or the corresponding shift vector must satisfy
    \begin{equation}
        (v_L,v_R ) \notin I_\theta^{*}\,.
    \end{equation}
so that there is no $(p_L,p_R)$ that cancels both terms in \eqref{eq:Hvec} and \eqref{eq:Hspinor}.

\section{Representation theory problem}\label{sec:rep_th}

We want to approach the search for orbifolds realizing our cancellation mechanism from a group theoretic point of view. This will follow closely the approach in \cite{GrootNibbelink:2017luf,Larotonda:2026hxy}, adapting the appropriate tools to the representation theory of non-abelian groups.

Recall that we are looking for partition functions in which each term in the sum \eqref{eq:orbifold_partition_function} vanishes independently. In the toroidal setting, it can be proven (see Appendix B of \cite{Larotonda:2026hxy}) that the untwisted sector fermionic partition function
\begin{equation}
    \mathcal{Z}^F[1,g] = \mathcal{Z}^{NS}[1,g] - \mathcal{Z}^R[1,g]\,,
\end{equation}
vanishes if and only if the element $g$ preserves a supercharge.

This can be explicitly checked by the following.
In light cone quantization of 10D Type II string theory, the 16 manifest supercharges transform in the $(\mathbf{8},\mathbf{1})\oplus (\mathbf{1},\mathbf{8})$ representation of $Spin(8)_L\times Spin(8)_R$. A $T^6$ compactification breaks $Spin(8)$ to $\Gamma \times Spin(2)$, where $Spin(2)$ is the massless little group in four dimensions and $\Gamma$ embeds in $SU(4)\simeq Spin(6)$. Under the decomposition $Spin(8)\rightarrow SU(4) \times Spin(2)$, we have
\begin{equation}
    \mathbf{8}\rightarrow \mathbf{4}_{+1/2},
\end{equation}
so the supercharges transform on the complex $(\mathbf{4},\mathbf{1}) \oplus (\mathbf{1},\mathbf{4})$ of $SU(4)_L\times SU(4)_R$.

Therefore, to check if an element $g \in G$ preserves a supercharge, we will construct the explicit representation $\rho_F:G\mapsto SU(4)_L\times SU(4)_R$ of its action on fermions, and check if it has any eigenvalue 1.

If $\mathcal{Z}[1,g]=0$, then by modular covariance the entire orbit vanishes:
\begin{equation}\label{eq:modularorbitZ}
    \mathcal{Z}^F[g^a,g^b] = 0\,.
\end{equation}
We will then build models in which the sum \eqref{eq:orbifold_partition_function} is given essentially only by a sum of terms like \eqref{eq:modularorbitZ} for elements which preserve a supercharge.
Crucially, for SUSY to be broken in the full theory, there can be no charge that is preserved by all elements simultaneously.
This means that $\rho_F$ cannot have any trivial subrepresentation.

In principle there can be many other commuting pairs that are not of the form $(g^a,g^b)$, but in most of the groups we need to examine there will be relatively few exceptions. It is therefore convenient algorithmically to impose the conditions we are about to spell out first, and then check by hand if there are any commuting pairs which would give a non-vanishing contribution~$\mathcal{Z}[f,g]\neq 0$.\footnote{Roughly, for a generic group of order $|G|=n$, imposing a condition of each element scales as $O(n)$, while imposing conditions on each pair of elements scales as $O(n^2)$.}

The conditions are summarized as follows: we are looking for a faithful representation $\rho_F:G \rightarrow GL(V)$ of a finite group $G$ with the following properties:
\begin{itemize}
    \item[] \textbf{I.} $V$ is 8-dimensional, and $\rho_F$ splits as the direct sum of two 4D subrepresentations: the left and right movers.
    \item[] \textbf{II.} In each 4D block, every element has determinant 1, to land in $SU(4)_L\times SU(4)_R$.
    \item[] \textbf{III.} Every element fixes some vector, i.e.~$\forall g \in G, \, \exists \, v \in V$ such that $\rho_F(g) \cdot v = v \,$.
    Equivalently, every matrix $\rho_F(g)$ has an eigenvalue 1.
    \item[] \textbf{IV.} $\rho_F$ does not have a trivial subrepresentation, i.e.~no vector is fixed by the whole $G$.
    \item[] \textbf{V.} The induced $SO(6)_L\times SO(6)_R$ representation $\rho_B$ on bosons can be obtained as a crystallographic symmetry of a 6D lattice.
\end{itemize}

For abelian groups, it was shown in \cite{Larotonda:2026hxy} that conditions \textbf{I.}-\textbf{IV.} are already strong enough to bound the order of the group, which led to a full classification of point groups. Condition \textbf{V.}, which is equivalent to a restriction to toroidal orbifold, imposed additional restrictions but was not crucial for the finiteness argument.

For non-abelian groups things are more complicated in this sense: already in \cite{Larotonda:2026hxy} representations satisfying \textbf{I.}-\textbf{IV.} were given for dihedral group $D_{2n}$ of arbitrarily high order. We therefore really have to rely on the classification of crystallographic symmetries to provide a full classification. This necessarily restricts us to the setting of toroidal orbifolds.

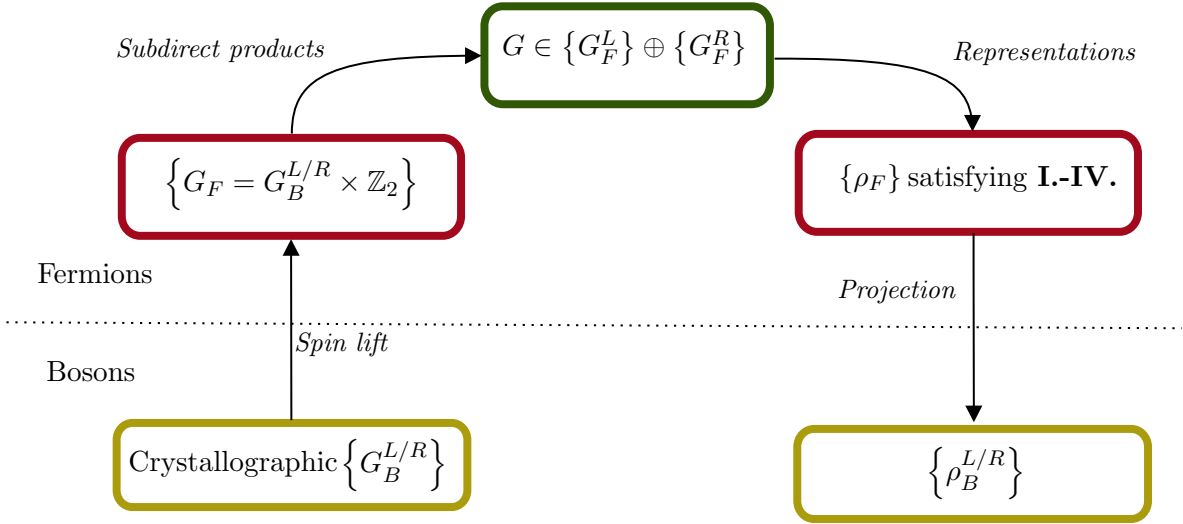
\begin{figure}

\tikzset{every picture/.style={line width=0.75pt}} 

\begin{tikzpicture}[x=0.70pt,y=0.75pt,yscale=-1,xscale=1]

\draw  [color={rgb, 255:red, 48; green, 88; blue, 5 }  ,draw opacity=1 ][fill={rgb, 255:red, 184; green, 233; blue, 134 }  ,fill opacity=0 ][line width=3]  (274,305.31) .. controls (274,299.89) and (278.4,295.49) .. (283.83,295.49) -- (417.69,295.49) .. controls (423.12,295.49) and (427.52,299.89) .. (427.52,305.31) -- (427.52,334.79) .. controls (427.52,340.22) and (423.12,344.62) .. (417.69,344.62) -- (283.83,344.62) .. controls (278.4,344.62) and (274,340.22) .. (274,334.79) -- cycle ;
\draw  [color={rgb, 255:red, 170; green, 156; blue, 12 }  ,draw opacity=1 ][fill={rgb, 255:red, 184; green, 233; blue, 134 }  ,fill opacity=0 ][line width=3]  (76.23,512.93) .. controls (76.23,508.27) and (80.01,504.49) .. (84.67,504.49) -- (256.39,504.49) .. controls (261.05,504.49) and (264.83,508.27) .. (264.83,512.93) -- (264.83,538.24) .. controls (264.83,542.9) and (261.05,546.68) .. (256.39,546.68) -- (84.67,546.68) .. controls (80.01,546.68) and (76.23,542.9) .. (76.23,538.24) -- cycle ;
\draw    (170.44,359.32) .. controls (170.52,329.56) and (193.06,320.44) .. (270.16,320.26) ;
\draw [shift={(272.52,320.26)}, rotate = 180] [fill={rgb, 255:red, 0; green, 0; blue, 0 }  ][line width=0.08]  [draw opacity=0] (8.93,-4.29) -- (0,0) -- (8.93,4.29) -- cycle    ;
\draw    (170.19,503.13) -- (170.04,415.29) ;
\draw [shift={(170.03,412.29)}, rotate = 89.9] [fill={rgb, 255:red, 0; green, 0; blue, 0 }  ][line width=0.08]  [draw opacity=0] (8.93,-4.29) -- (0,0) -- (8.93,4.29) -- cycle    ;
\draw  [color={rgb, 255:red, 163; green, 9; blue, 29 }  ,draw opacity=1 ][fill={rgb, 255:red, 219; green, 81; blue, 21 }  ,fill opacity=0 ][line width=3]  (79,371.31) .. controls (79,365.89) and (83.4,361.49) .. (88.83,361.49) -- (251.61,361.49) .. controls (257.04,361.49) and (261.44,365.89) .. (261.44,371.31) -- (261.44,400.79) .. controls (261.44,406.22) and (257.04,410.62) .. (251.61,410.62) -- (88.83,410.62) .. controls (83.4,410.62) and (79,406.22) .. (79,400.79) -- cycle ;
\draw  [color={rgb, 255:red, 163; green, 9; blue, 29 }  ,draw opacity=1 ][fill={rgb, 255:red, 219; green, 81; blue, 21 }  ,fill opacity=0 ][line width=3]  (443,369.31) .. controls (443,363.89) and (447.4,359.49) .. (452.83,359.49) -- (615.61,359.49) .. controls (621.04,359.49) and (625.44,363.89) .. (625.44,369.31) -- (625.44,398.79) .. controls (625.44,404.22) and (621.04,408.62) .. (615.61,408.62) -- (452.83,408.62) .. controls (447.4,408.62) and (443,404.22) .. (443,398.79) -- cycle ;
\draw    (536.35,355.18) .. controls (534.85,330.66) and (513.98,321.45) .. (430.03,321.75) ;
\draw [shift={(536.44,358.32)}, rotate = 270.08] [fill={rgb, 255:red, 0; green, 0; blue, 0 }  ][line width=0.08]  [draw opacity=0] (8.93,-4.29) -- (0,0) -- (8.93,4.29) -- cycle    ;
\draw    (537.18,501.13) -- (537.03,429.13) -- (537.03,409.29) ;
\draw [shift={(537.19,504.13)}, rotate = 269.88] [fill={rgb, 255:red, 0; green, 0; blue, 0 }  ][line width=0.08]  [draw opacity=0] (8.93,-4.29) -- (0,0) -- (8.93,4.29) -- cycle    ;
\draw  [dash pattern={on 0.84pt off 2.51pt}]  (17,454) -- (649,457) ;
\draw  [color={rgb, 255:red, 170; green, 156; blue, 12 }  ,draw opacity=1 ][fill={rgb, 255:red, 184; green, 233; blue, 134 }  ,fill opacity=0 ][line width=3]  (442.23,516.53) .. controls (442.23,511.54) and (446.27,507.49) .. (451.27,507.49) -- (621.79,507.49) .. controls (626.78,507.49) and (630.83,511.54) .. (630.83,516.53) -- (630.83,543.64) .. controls (630.83,548.63) and (626.78,552.68) .. (621.79,552.68) -- (451.27,552.68) .. controls (446.27,552.68) and (442.23,548.63) .. (442.23,543.64) -- cycle ;

\draw (81,517) node [anchor=north west][inner sep=0.75pt]   [align=left] {Crystallographic};
\draw (195,511) node [anchor=north west][inner sep=0.75pt]    {$\left\{G_{B}^{L/R}\right\}$};
\draw (170.11,457.71) node [anchor=north west][inner sep=0.75pt]   [align=left] {\textit{{\small Spin lift}}};
\draw (100,370) node [anchor=north west][inner sep=0.75pt]    {$\left\{G_{F} =G_{B}^{L/R} \times \mathbb{Z}_{2}\right\}$};
\draw (463,373) node [anchor=north west][inner sep=0.75pt]    {$\{\rho _{F}\}$};
\draw (74,310) node [anchor=north west][inner sep=0.75pt]   [align=left] {\textit{{\small Subdirect products}}};
\draw (282,306) node [anchor=north west][inner sep=0.75pt]    {$G\in \left\{G_{F}^{L}\right\} \oplus \left\{G_{F}^{R}\right\}$};
\draw (500,374) node [anchor=north west][inner sep=0.75pt]   [align=left] {satisfying \textbf{I.-IV.}};
\draw (524,312) node [anchor=north west][inner sep=0.75pt]   [align=left] {\textit{{\small Representations}}};
\draw (462,431) node [anchor=north west][inner sep=0.75pt]   [align=left] {{\small \textit{Projection}}};
\draw (510,515) node [anchor=north west][inner sep=0.75pt]    {$\left\{\rho _{B}^{L/R}\right\}$};
\draw (37,472) node [anchor=north west][inner sep=0.75pt]   [align=left] {Bosons};
\draw (32,423) node [anchor=north west][inner sep=0.75pt]   [align=left] {Fermions};
\end{tikzpicture}
    \caption{Schematic representation of the algorithm used to classify the orbifold groups. We start from candidate crystallographic bosonic groups $G_B^{L/R}$, lift them to the fermionic group $G_F$ and combine left and right sectors via all possible subdirect products. We then identify fermionic representations $\rho_F$ satisfying our constraints \textbf{I.}-\textbf{IV.}. Finally, we project back to the bosonic actions and verify that the induced representations $\rho_B^{L/R}$ are still crystallographic.}
    \label{fig:argument_flow}
\end{figure}

\subsection{Strategy for the classification}

Let us now explain the strategy we adopted. We will be more explicit about the technical details when presenting the full $T^5\times S^1$ classification. First of all, one can note that the list of crystallographic symmetries in each dimension is finite. One could in principle compute all possible combinations of bosonic crystallographic representations $\rho_B^{L/R}$, compute their spin lifts, and only verify that they satisfy conditions \textbf{I.}-\textbf{IV.}, since \textbf{V.} is guaranteed. Apart from being computationally intense, it relies on algorithmically computing spin lifts of arbitrary $SO(6)$ subgroups, which is not straightforward. We will get around this issue by mostly working with the fermionic representations themselves.

For most of the group theoretic calculations we use the software GAP \cite{GAP4}, while we access the data on crystallographic symmetries through the package CARAT \cite{Plesken:zm0025}.

The first piece of information we can get from the crystallographic classification in CARAT is the list of abstract groups, acting on bosons, $G_B^{L/R}$, which can be considered as candidates. In dimension $d$, these are obtained as subgroups of a relatively short list of maximal subgroups of $GL(d;\Z)$. We will explore the $d=5$ case in detail, leaving the $d=6$ classification for future work.

The corresponding group $G_F^{L/R}$ acting on fermions is a priori an extension:
\begin{equation}
    0 \longrightarrow \Z_2 \longrightarrow G_F^{L/R} \longrightarrow G_B^{L/R} \longrightarrow 0\,.
\end{equation}
Anomaly cancellation imposes that a certain topological quantity associated to the bosonic representation $\rho_B:G_B\rightarrow SO(6)$ vanishes \cite{Freed:1987qk}.
This is the second Stiefel-Whitney class $w_2(\rho_B)$ of the representation, which can be given an interpretation in the general bordism framework \cite{Delgado:2026qvy}. We expand on this in \hyperref[sec:anomalies]{Section 4}. For us, the main implication is, as explained in \hyperref[app:ext]{Appendix B}, that the extension is trivial
\begin{equation}
    w_2(\rho_B^{L/R})=0 \quad \Longrightarrow \quad G_F^{L/R} \simeq G_B^{L/R} \times \Z_2\,.
\end{equation}
Leveraging this, what we can do is consider the list of crystallographic groups in $d$ dimensions, and obtain the corresponding list of fermionic groups acting on the left/right:
\begin{equation}
    G_F^{L/R} \in \left\{ G_B^{L/R} \times \Z_2 \,\bigg{|} \, G_B ^{L/R}\text{ subgroup of }GL(d;\Z) \right\}\,.
\end{equation}
An additional freedom of asymmetric orbifolds is that the bosonic group acting on the left and on the right can in principle be different, as long as they both are symmetries of the same lattice.
Concretely, we can take generators $\{g_i^{L/R}\}$ of some possibly different five-dimensional crystallographic groups on the left and on the right, and combine them by taking direct sums. This means that for some arbitrary sets of generators $\{g_i^{L/R}\}$, the full bosonic group is generated by
\begin{equation}
    g_i :=g_i^L \oplus g_i^R\,.
\end{equation}
As explained in \hyperref[app:subdirect]{Appendix C}, a group generated by these kind of elements is called a subdirect product of $G_B^L$ and $G_B^R$. Consequently, the full fermionic group is a subdirect product of $G_F^L$ and $G_F^R$. 

The full list of candidate groups we consider is then
\begin{equation}
    G_F \in \left\{ \text{subdirect products of $G_B^{L} \times \Z_2,G_B^{R} \times \Z_2$ } \bigg{|} \, G_B ^{L/R}\text{ subgroup of }GL(d;\Z)  \right\}\,.
\end{equation}
For all $G_F$'s in this big list, we construct all representations $\rho_F:G_F\rightarrow SU(4)_L\times SU(4)_R$ satisfying conditions \textbf{I.}-\textbf{IV.} above, using the software GAP.

So far all of this has been at the abstract group level, we have not yet looked at crystallographic representations per se. To make the difference clear, take the example of $G= \Z_5$. There are lattices in $d\geq 4$ which admit a symmetry of order 5, but not with every real representation of $\Z_5$. From the corresponding cyclotomic polynomial, see \hyperref[app:cyclotomic]{Appendix A}, one can see that the representation sending the generator to a rotation with eigenvalues $(e^{2\pi i/5},e^{-2\pi i/5},e^{2\pi i/5},e^{-2\pi i/5})$ can never be crystallographic, i.e.~it can never be conjugate to a $GL(5,\Z)$ matrix.

We therefore have to ensure that, given a good fermionic representation $\rho_F$, the induced representations $\rho_B^{L/R}$ are crystallographic and, crucially, they are symmetries \textit{of the same lattice}.

In order to do so, for every lattice in $5$ dimensions, we generated through CARAT a full list of its abstract symmetry subgroups and their representations. If then both $\rho_B^L$ and $\rho_B^R$ appear in the list associated to the same crystal, they can be simultaneously represented as integral matrices that are automorphisms of the same lattice, and the group $G$ realizes a toroidal orbifold.

We now have to recall that so far we have imposed a slightly weaker condition than the one we need to ensure that the one-loop vacuum energy vanishes. Condition \textbf{III.} above guarantees the vanishing of all sectors \eqref{eq:modularorbitZ}, but there may be other non-trivial commuting pairs entering the orbifold sum \eqref{eq:orbifold_partition_function}. For most of our solutions there are relatively few such pairs, sometimes none at all for groups of lower order. We therefore find it convenient to analyse them separately.

For each representation that satisfies all properties above, we impose the additional condition to ensure vanishing of the cosmological constant:
\begin{itemize}
    \item[] \textbf{VI.} For every commuting pair of elements $(f,g)$, they fix a common vector, i.e.~$\exists\, v \in V$ such that $\rho(f)\cdot v = v = \rho(g) \cdot v $
\end{itemize}

Models obtained this way are not necessarily consistent, since they may suffer from global anomalies, breaking modular invariance \cite{Freed:1987qk}. 

 We impose anomaly cancellation in steps, according to the logic in \hyperref[sec:as_orbifolds]{Section 2}, abelian anomaly cancellation amounts to checking level matching for all elements, which can be done algorithmically through \eqref{eq:modInv}. The extra mod 2 condition in \cite{Narain:1986qm} for even order groups is equivalent to checking that our specific representation has $w_2(\rho_{B}^{L/R})=0$, see \cite{Delgado:2026qvy}. This condition, as discussed above, is also straightforward to check algorithmically: we have to require that both on the left and on the right the \textit{faithfully realized} fermionic group $\widetilde{G}_F^{L/R}$ is a trivial extension $\widetilde{G}_B^{L/R} \times \Z_2$ of the \textit{faithfully realized} bosonic group $\widetilde{G}_B^{L/R}$. 

For non-abelian groups for which anomalies are not fully detected by abelian subgroups, we will check the extra conditions case by case, as explained in \hyperref[sec:anomalies]{Section 4}.

Interestingly, all solutions to properties \textbf{I.}-\textbf{V.} satisfy anomaly cancellation, even on the purely non-abelian anomalies, so we could not exclude any model this way. This cancellation was, however, not trivial to expect from the start. We will comment on this coincidence again later, after we have introduced the proper tools to study anomalies.

Finally, we examine the twisted sectors to make sure we have not accidentally restored SUSY. As discussed above, if there are point group elements acting trivially on bosons either on the left and on the right they can generate massless gravitini in their twisted sector. We will examine individually the solutions that have passed all other checks so far, and if needed manually add shifts to the generators to lift the gravitino masses through \eqref{eq:Hvec}, \eqref{eq:Hspinor}.

\section{Anomalies for non-abelian groups}\label{sec:anomalies}

In this Section we want to study global anomalies for our models. In particular, since level matching is straightforward to check algorithmically through \eqref{eq:levelMatch}, we want to know for which groups there can be additional anomalies in modular invariance for level matched models, and compute this extra anomaly in such cases.

After a modular transformation, the partition function could pick up a phase:
\begin{equation}\label{eq:anomalygeneral}
    \mathcal{Z}(\tau,\bar{\tau}) \mapsto\mathcal{Z}\left(\frac{a\tau + b}{c\tau + d},\frac{a\bar{\tau} + b}{c\bar{\tau} + d}\right)= \mathcal{Z}(\tau,\bar{\tau})e^{2 \pi i \mathcal{A}}
\end{equation}
Note that since the anomaly $\mathcal{A}$ always appears as a phase with the normalization as in \eqref{eq:anomalygeneral}, all equations involving it have to be taken up to integers, i.e.~as valued in $\R/\Z$.

There are multiple ways to approach the computation of this phase. Historically, \cite{Freed:1987qk} provided a sufficient condition for the vanishing of these anomalies, in terms of the vanishing of some cohomology classes
\begin{equation}\label{eq:cohomology_conditions}
    w_2(\rho)=0 \in H^2(G;\Z_2) \quad \text{ and } \quad \frac{1}{2}p_1(\rho)=0 \in H^4(G;\Z)\,,
\end{equation}associated with the representations our fields transform in\footnote{We review the definition of characteristic classes of representations in \hyperref[app:char]{Appendix B}. The fact that $\tfrac{1}{2}p_1(\rho)$ is well defined as an integral class is a consequence of $w_2(\rho)=0$, see e.g.~\cite{Freed:1987qk}.}.
For abelian groups, this is in fact equivalent to imposing level matching \cite{Freed:1987qk}. The computation of the Pontryagin class $p_1$ is however difficult for general groups if attempted directly in group cohomology.

From a more modern point of view, we know that gauge anomalies of fermions are computed as $\eta$-invariants of the Dirac operator $\mathcal{D}_{\rho_F}$ coupled to the corresponding representation $\rho_F$, see e.g.~\cite{Garcia-Etxebarria:2018ajm}.
The $\eta$-invariant is defined as a regularized sum on the eigenvalues of the Dirac operator
\begin{equation}
    \eta(\rho) = \eta(\mathcal{D_\rho)} = \lim_{s\rightarrow 0} \sum_{\lambda \neq0} \frac{\text{sign}(\lambda)}{|\lambda|^s} \,.
\end{equation}
For discrete groups, these quantities compute gauge anomalies, and are in fact bordism invariants\footnote{In fact, the correct statement in Type II theories is that the most general anomaly is an element of
\begin{equation}
    \text{Hom}(\Omega_3^\Spin(BG \times B \Z_2^{F});U(1))\,.
\end{equation}
Bordism groups for products of spaces always split as
\begin{equation}
    \Omega_3^\Spin(BG\times B\Z_2^{F}) \simeq \Omega_3^\Spin(BG) \oplus \Omega_3^\Spin(B\Z_2^{F}) \oplus \Omega_3^\Spin(BG \wedge B\Z_2^{F})\,.
\end{equation}
In our models, vanishing of the anomalies associated to the last two summands is covered by the cohomological conditions $w_1(\rho)=0= w_2(\rho)$\cite{Delgado:2026qvy}, which we are guaranteed to satisfy because of the extension properties discussed in \hyperref[app:ext]{Appendix B}. We therefore from now on focus on the pure gauge anomalies \eqref{eq:global_anomalies}.} 
\begin{equation}\label{eq:global_anomalies}
    \mathcal{A}_\text{gauge} \in \text{Hom}(\Omega_3^\Spin(BG);U(1))\,,
\end{equation}
and computing these bordism groups can give a lot of information on the possible anomalies.

The equivalence with level matching in the abelian case has been recently explored in detail in \cite{Cheng:2026abk}. The relation between the first condition in \eqref{eq:cohomology_conditions} has also recently been proven to be equivalent to the vanishing of the mixed anomaly of $G$ with the GSO projection \cite{Delgado:2026qvy}.

For non-abelian groups, the equivalence of \eqref{eq:cohomology_conditions} with the vanishing of \eqref{eq:global_anomalies} is still valid: in fact the calculations in \cite{Freed:1987qk} are nothing but the calculation of the $\eta$-invariant of the representation in which bosons transform on a specific class of 3-manifolds, namely mapping tori for modular transformations.

We take here the point of view of \cite{Garcia-Etxebarria:2018ajm}, wherein it is argued that more general anomalies should be considered, where the topology is allowed to change ``along the mapping torus", leading to the requirement of the vanishing of the anomaly theory on all bordism generators.

This language gives us a way to determine a priori if a given non-abelian group $G$ can give rise to additional anomalies beyond those fixed by level matching, as follows.
Consider a non-abelian group $G$, and some abelian subgroup $A \subset G$, then the inclusion
\begin{equation}
    i:A \rightarrow G
\end{equation}
induces a map on the corresponding bordism group\footnote{More generally, $X\mapsto\Omega_d^{\xi}(X)$ is a covariant functor from the category of topological spaces to that of abelian groups. This, together with the induced map from $A\rightarrow G$ on the corresponding classifying spaces $BA\rightarrow BG$, gives the map above.}:
\begin{equation}
    i_* : \Omega_3^\Spin(BA)\rightarrow \Omega_3^\Spin(BG)\,.
\end{equation}
If all elements in $\Omega_3^\Spin(BG)$ lie in the image of such a map for some abelian subgroup $A$, then all the anomalies of any two-dimensional theory with discrete symmetry $G$ are induced from those of its abelian subgroups. For groups $G$ for which this is the case, see e.g.~\cite{Larotonda:2026hxy} for some non-abelian examples, and \cite{Cheng:2026abk} for the proof of this equivalence for cyclic groups, the level matching condition \eqref{eq:modInv} is then sufficient to ensure modular invariance.

The bordism results in \cite{Davighi:2022icj,Braeger:2025kra,Larotonda:2026hxy} imply that this is the case for $G=S_3 \times \Z_k$: there are no anomalies apart from those induced by abelian subgroups. Therefore level-matched $S_3 \times \Z_k$ models are guaranteed to be modular invariant.

We find this to also be the case for some more groups, namely  $D_{12}$ and $Dic_{12}$ and products of these with cyclic factors $\Z_k$.
For all other groups in our list of solutions we are going to have to compute the $\eta$-invariants that generate \eqref{eq:global_anomalies} explicitly.

In 2D the situation 
is slightly more subtle, because of the presence of pure gravitational anomalies \cite{Alvarez-Gaume:1983ihn}, which are obstructions to the $\eta$-invariant being a $\Omega_3^\Spin(BG)$-invariant. The gauge anomaly of a fermion transforming in a representation $\rho$ of $G$ is in fact given as the difference of the $\eta$-invariant of $\rho$ and the $\eta$-invariant of an uncharged fermion, see e.g.~\cite{Garcia-Etxebarria:2018ajm,Basile:2023zng,Dierigl:2025rfn} for applications in the context of cyclic groups.

The quantity that we then want to compute is the reduced $\eta$-invariant
\begin{equation}
    \widetilde{\eta}(\rho,X):=
    \eta(\rho,X)- \text{dim}(\rho) \eta(\textbf{1},X) \quad\in \quad\text{Hom}(\Omega_3^\Spin(BG),U(1))\,,
\end{equation}
for a complete set of generators $X$ of the corresponding bordism group $\Omega_3^\Spin(BG)$.
We will do it case by case by decomposing $\rho$ into irreducible representations and using that $\eta$-invariants are additive in their first entry:
\begin{equation}
    \eta(\rho_1\oplus\rho_2,X) = \eta(\rho_1,X) + \eta(\rho_2,X)\,.
\end{equation}

As for the choice of bordism generators $X$, for all groups relevant to our classification they turn out to be so-called spherical space forms \cite{Gilkey_book}. These are quotients of $S^3$ obtained as follows.

Since $SU(2) \simeq S^3$, all finite proper subgroups $\Gamma$ of $SU(2)$ act freely on $S^3$, and can be used to build smooth manifolds 
\begin{equation}
    X_\Gamma = S^3/\Gamma\,.
\end{equation}
If we have a faithful representation $\rho:G\rightarrow SU(2)$, we can quotient by its image in $SU(2)$ and obtain a manifold
\begin{equation}\label{eq:sphericalspace}
    X_\rho := S^3/\rho(G)\,.
\end{equation}
For these spaces, a concrete formula to compute $\eta$-invariants is known \cite{Gilkey_book,Garcia-Etxebarria:2018ajm}. 
The eta-invariant of a fermion transforming in the representation $\rho_F$ on a spherical space form, defined as \eqref{eq:sphericalspace} by a representation $\rho_\text{space}$, is given by
\begin{equation} \label{eq:etaInvariants}
    \eta(\rho_F,X_{\rho_\text{space}}) = \frac{1}{|G|} \sum_{g\neq 1} \frac{\text{Tr}(\rho_F(g))}{\text{Det}(\rho_\text{space}(g)-\mathbb{I}_2)}\,.
\end{equation}

Notice that so far we have only talked about fermions. These are not the only potentially anomalous degrees of freedom living on the string worldsheet. In two dimensions, chiral bosons can carry anomalies. In the symmetric orbifold literature this is usually ignored, since the left- and right-moving coordinates are acted upon in the same way by the orbifold, and thus they combine to a non-chiral representation of the orbifold group $G$.

Since we deal with asymmetric orbifolds, $G$ acts differently on left and right movers, so their potentially anomalous representations do not automatically cancel with each other, and have to be examined carefully.

The main complication is that chiral bosons are not only acted upon by the point group, but rather by the full space groups, unlike worldsheet fermions which are not sensitive to shifts.
Thankfully, for the concrete examples that appear in our classification, we do not have to consider a general action, and we are allowed to limit ourselves to orbifold actions which are either purely rotation, or purely shift.

Let us start by letting all shifts in the orbifold group be trivial, i.e.~only considering the action of the point group $P_G$. Then the chiral bosons transform in the linear $SO(6)$ representation $\rho_B$, and one can apply the tools of \cite{Monnier:2011mv,Monnier:2013kna,Hsieh:2020jpj} to write the anomaly contribution of bosons as
\begin{equation}\label{eq:bosoniceta}
    \mathcal{A}^B_{\text{twist}}= -\frac{1}{2}\eta(\rho_B^\C)\,.
\end{equation}

An $\eta$-invariant of a Dirac operator appears here despite the degrees of freedom being bosonic. This should not be surprising, since the theory of a free chiral boson is equivalent through fermionization to that of a free complex fermion, see e.g.~\cite{DiFrancesco:1997nk, Blumenhagen:2009zz}. Furthermore, in general, defining the theory of self-dual $p$-forms in $2p+2$ dimension requires specifying an additional geometric datum: a quadratic refinement of the differential cohomology pairing \cite{Witten:1996hc,Belov:2006jd}. In 2D a spin structure provides a natural choice for such a refinement, which is precisely \eqref{eq:bosoniceta} \cite{Hsieh:2020jpj}.

The prefactor also deserves some explanation: the factor of $\tfrac{1}{2}$ is because the representation we are examining is real, and the $\eta$-invariant is defined for a complex representation. So what one does is complexifying as $\rho_B^\C = \rho_B \otimes \C$, computing the $\eta$-invariant of this representation, and divide by two since inside $\rho_B^\C$ there are two $\C$-conjugate copies of $\rho_B$, which are equivalent from this point of view. The minus sign has to be taken as relative to a fermion of the same chirality transforming in the same linear representation.\footnote{The idea behind the minus sign is that a fermionic partition function is roughly $$\mathcal{Z}_F \sim \sqrt{\text{Det}(i \mathcal{D})} \,,$$ while a chiral boson can be written as $$\mathcal{Z}_B \sim \frac{1}{\sqrt{\text{Det}(i\mathcal{D})}}\,,$$ for an appropriately defined Dirac operator. Their respective anomalous phases are then conjugate to each other. See e.g.~\cite{Monnier:2011mv,Monnier:2013kna} for the precise definition of the bundles of which $\mathcal{Z}_{F/B}$ are sections of, and the formal proof of these statements.}

Going back to a space group which contains shifts, for the cases at hand we are allowed to simplify things further. For us, the necessity to add asymmetric shifts is to get rid of massless twisted gravitini should they not be already projected out by the point group action. It so happens that in our list of models, this is necessary only for the point group $P_G = S_3$, for which we have already seen that there are no purely non-abelian anomalies. This means that level matching is enough to guarantee modular invariance. Adding shifts without modifying the group structure will not spoil this, so the contribution to the anomaly is precisely what appears in \eqref{eq:modInv}.
\\In the language of this Section, we can rewrite the contribution to level matching of an order $k$ shift as a contribution to the corresponding anomaly
\begin{equation}
    \mathcal{A}_\text{shift}^B = k\frac{(v^*)^2}{2}
\end{equation}
on an order $k$ generator in $\Omega_3^\Spin(B\Z_k)$, which can always be taken as a three-dimensional lens space $L_k^3$ with an order $k$ holonomy turned on.

In total, we want the full anomaly theory to vanish, so we demand
\begin{equation}
    \mathcal{A}^F_L[X] + \mathcal{A}^B_L[X] - \mathcal{A}^F_R[X] - \mathcal{A}^B_R[X] = 0
\end{equation}
for a full set of generators $X$ of the corresponding bordism group.

\subsection{\texorpdfstring{$Q_8$}{Q8}}\label{sec:Q8_anomaly}

For the quaternion group, abelian subgroups are not enough to describe the full anomaly. In fact we have \cite{Davighi:2022icj}:
\begin{equation}\label{eq:q8bord}
    \Omega_3^\Spin(BQ_8) \simeq \Z_8 \oplus \Z_4^{\oplus2}\,,
\end{equation}
which is strictly bigger than the bordism associated to its maximal abelian subgroups $\Z_4$ \cite{Guo:2018vij}:
\begin{equation}
    \Omega_3^\Spin(B\Z_4) \simeq \Z_8 \oplus \Z_2\,.
\end{equation}
The generators of \eqref{eq:q8bord} are known, and so are all $\eta$-invariants computed on them \cite{Botvinnik:1995}. Let us describe them briefly.

We use the standard presentation of $Q_8$ as
\begin{equation}
    Q_8 = \langle i, j, k \mid i^2 = j^2 = k^2 = ijk \rangle\,.
\end{equation}
Borrowing notation from \cite{Botvinnik:1995}, we denote the cyclic $\Z_4$ subgroups of $Q_8$ as
\begin{equation}
    H_i = \langle i \rangle, \quad H_j = \langle j \rangle, \quad H_k = \langle k \rangle\,,
\end{equation}
and the faithful representation of $Q_8$ as $\tau$:
\begin{equation}\label{eq:taurep}
    \tau(i) = \begin{pmatrix} i & 0 \\ 0 & -i \end{pmatrix}, \quad \tau(j) = \begin{pmatrix} 0 & 1 \\ -1 & 0 \end{pmatrix}, \quad \tau(k) = \begin{pmatrix} 0 & i \\ i & 0 \end{pmatrix}.
\end{equation}
The three non-trivial 1D representations are denoted by the element they represent trivially, as follows:
\begin{equation}
    \begin{aligned}
        \rho_i(i) &= 1,& \quad \rho_i(j) &= -1,& \quad \rho_i(k) &= -1, \\
        \rho_j(i) &= -1,& \quad \rho_j(j) &= 1, &\quad \rho_j(k) &= -1, \\
        \rho_k(i) &= -1,& \quad \rho_k(j) &= -1,& \quad \rho_k(k) &= 1.
    \end{aligned}
\end{equation}
\\Then the order 8 generator of $\Omega_3^\Spin(BQ_8)$ is given by the spherical space form
\begin{equation}
    X_{Q_8} = S^3/\tau(Q_8) \,,
\end{equation}
as can be seen by the corresponding $\eta$-invariants
\begin{equation}
    \begin{aligned}
        \eta(\rho_i,X_{Q_8}) = \eta(\rho_j,X_{Q_8})&= \eta(\rho_k,X_{Q_8})=\frac{1}{2}\,, \\
        \eta(\rho_\tau,X_{Q_8}) &= \frac{7}{8}\,.
    \end{aligned}
\end{equation}
The $\Z_4\oplus \Z_4$ summand in \eqref{eq:q8bord} is generated by choosing any two out of
\begin{equation}\label{eq:Q8BORD_Z4_generators}
    X_i = S^3/\tau(H_i), \quad X_j = S^3/\tau(H_j), \quad X_k = S^3/\tau(H_k).
\end{equation}
And the corresponding $\eta$-invariants read as follows:
\begin{equation}
    \begin{aligned}
        \eta(\rho_\tau,X_i) &= \eta(\rho_\tau,X_j) = \eta(\rho_\tau,X_k)=\frac{3}{4}\,, \\
        \eta(\rho_g,X_h) &= \frac{1}{2}(1-\delta_{g,h}) \quad \text{for }g,h \in \{i,j,k \}\,.
    \end{aligned}
\end{equation}

\subsection{\texorpdfstring{$Dic_{12}$}{Dic12}}\label{sec:Dic_12_anomaly}

For the dicyclic group of order 12, we use a presentation
\begin{equation}
    Dic_{12} = \langle r,s \big{|} r^6 = s^4 = 1, r^3 = s^2, srs^{-1}=r^{-1} \rangle\,.
\end{equation}
We begin with a computation of the relevant bordism group. To this end, we use the Atiyah-Hirzebruch spectral sequence (from now on AHSS \cite{atiyah1962vector}, see e.g.~\cite{Garcia-Etxebarria:2018ajm,Davighi:2022icj} for physics applications in a similar context) for the fibration $pt\rightarrow Dic_{12} \rightarrow Dic_{12}$, whose second page is defined as
\begin{equation}
    E^2_{p,q} = H_p(BG; \Omega_{q}^\Spin(pt))\,.
\end{equation}
The integral homology of $Dic_{12}$ reads in low degree \cite{brown1982cohomology}
\begin{equation}
    H_1(Dic_{12})\simeq \Z_4, \quad H_2(Dic_{12})\simeq0, \quad H_3(Dic_{12})\simeq \Z_{12} \simeq \Z_3 \oplus \Z_4\,,
\end{equation}
from which we build $E^2_{p,q}$ using the Universal Coefficient Theorem (UCT), obtaining:
\begin{equation}\label{eq:AHSSdic12}
    E^2_{p,q} : \quad
    \begin{array}{c|cccc}
        3 & \quad & & & \\
        2 & \quad \mathbb{Z}_2 & \quad\Z_2 &  \quad\Z_2& \quad\Z_2 \\
        1 & \quad\mathbb{Z}_2 & \quad\Z_2 & \quad\Z_2 & \quad\Z_2 \\
        0 & \quad\mathbb{Z} & \quad\Z_4 & \quad0 & \quad\Z_{4} \oplus \Z_3 \\
        \hline
         q/p & 0 & 1 & 2 & 3
    \end{array}
\end{equation}
This already gives us the result localized at prime 3: the only non-trivial contribution is $E_{3,0}^2\simeq \Z_3$ and there are no possible non-zero differentials acting on it. We conclude that $\Omega_3^\Spin(BDic_{12})\simeq\Z_3$ at prime 3. For the prime 2 contribution, for now we limit ourselves to use \eqref{eq:AHSSdic12} to give a bound on the maximum order of the prime 2 contribution, which is 16, if all of $E^2_{3,0} \simeq \Z_4$, $E^2_{2,1} \simeq \Z_2$ and $E^2_{1,2} \simeq \Z_2$ survive until the $E^\infty$ page. This is enough to determine the result, without computing differentials explicitly.

To get the result, we leverage the semidirect product structure of $Dic_{12} \simeq \Z_3 \rtimes \Z_4$, see \cite{Davighi:2022icj} for the same technique applied to $SL(2,3)$. This same trick will allow us to get the result in most cases of interest.

The idea is to use the diagram
\begin{equation}
    \begin{tikzcd}
        0 \arrow[r] & \mathbb{Z}_{3} \arrow[r,"i"] & Dic_{12} \arrow[r, "\pi"] & \mathbb{Z}_4 \arrow[r] & 0 \,,\\
        & & & \mathbb{Z}_4 \arrow[u, equal] \arrow[ul, "j", hook] & 
    \end{tikzcd}
\end{equation}
where the upper line is the short exact sequence associated to the semidirect product structure of $Dic_{12}$, and $j$ is the inclusion of $\Z_4$ as a subgroup.

We apply the bordism functor to the triangle subdiagram to obtain

\begin{equation}
   \begin{tikzcd}
         \Omega_3^\Spin(BDic_{12}) \arrow[r, "\pi_*"] & \Omega_3^\Spin(B\mathbb{Z}_4) & \\
         & \Omega_3^\Spin(B\mathbb{Z}_4) \arrow[u, equal] \arrow[ul, "j_*"] & 
    \end{tikzcd}\,.
\end{equation}
This tells us that $j_*$ is injective, since if $j_*(x) = j_*(y)$, we can apply $\pi_*$ to both and obtain
\begin{equation}\label{eq:j_*_injectivity}
    \pi_*(j_*(x)) = x = y = \pi_*(j_*(y))\,.
\end{equation}
This means that the bordism group
\begin{equation}
    \Omega_3^\Spin(B\mathbb{Z}_4) \simeq \Z_2 \oplus \Z_{8}\,,
\end{equation}
injects into $\Omega_3^\Spin(BDic_{12})$. Since it saturates the order bound derived above, we can conclude that $j_*$ is an isomorphism at prime 2, and
\begin{equation}
    \Omega_3^\Spin(BDic_{12}) \simeq \Z_3 \oplus \Z_2 \oplus \Z_8\,.
\end{equation}
The injection $j_*$ can also help us find generators: since the generators of $\Omega_3^\Spin(B\mathbb{Z}_4)$ generate the $\Z_2\oplus \Z_8$ factor inside $\Omega_3^\Spin(BDic_{12})$, we can have a full generating set by taking them together with any manifold representing the $\Z_3$ factor.
To find this last generator, we need the representation theory of $Dic_{12}$.

There are three 1D representations, apart from the trivial one, given by
\begin{equation}
    \rho_{11}:
    \begin{cases}
        \rho_{11}(r) = 1 \\
        \rho_{11}(s) = -1
    \end{cases}\,,
    \qquad \quad
    \rho_{12}:
    \begin{cases}
        \rho_{12}(r) = -1 \\
        \rho_{12}(s) = i
    \end{cases}\,,
    \qquad\quad
    \rho_{13}:
    \begin{cases}
        \rho_{13}(r) = -1 \\
        \rho_{13}(s) = -i
    \end{cases}\,.
\end{equation}
And two 2D representations: one is faithful and given by
\begin{equation}\label{eq:repdic12}
    \rho_{21}(r) = \left(
        \begin{array}{cc}
         -\phi  & 0 \\
         0 & -\phi^{-1} \\
        \end{array}
    \right)\,, \qquad \quad
    \rho_{21}(s) = \left(
        \begin{array}{cc}
         0  & -1 \\
         1 & 0 \\
        \end{array}
    \right)\,,
\end{equation}  
and the other factors through the $S_3$ subgroup as
\begin{equation}
    \rho_{22}(r) = \left(
        \begin{array}{cc}
         \phi  & 0 \\
         0 & \phi^{-1} \\
        \end{array}
    \right)\,, \qquad \quad
    \rho_{22}(s) = \left(
        \begin{array}{cc}
         0  & 1 \\
         1 & 0 \\
        \end{array}
    \right)\,,
\end{equation}  
with $\phi = e^{\frac{2\pi i}{3}}$ a third root of unity.

To build the last bordism generator, we consider the spherical space form associated with the faithful representation \eqref{eq:repdic12}, which embeds in $SU(2)$. The $\eta$-invariant corresponding to the same representation on this space is given through \eqref{eq:etaInvariants} as
\begin{equation}
    \widetilde{\eta}(\rho_{21},X_{\rho_{21}}) = -\frac{1}{12}\,.
\end{equation}
This means it generates a $\Z_{12} \simeq \Z_3 \oplus \Z_4$ subgroup inside $\Omega_3^\Spin(BDic_{12})$. In total, we conclude that a complete set of generators for $\Omega_3^\Spin(BDic_{12})$ is given by
\begin{equation}
    \Omega_3^\Spin(BDic_{12}) = \langle X_{\rho_2}, L^3_4, \widetilde{L}_4^3 \, \rangle\,,
\end{equation}
where $L^3_4, \widetilde{L}_4^3$ are three-dimensional lens spaces $S^3/\Z_4$ with the two inequivalent choices of spin structure: these are known to generate $\Omega_3^\Spin(B\Z_4)$ \cite{Guo:2018vij}.

\subsection{\texorpdfstring{$Dic_{24}$}{Dic24}}\label{sec:Dic_24_anomaly}

For the dicyclic group of order 24, we use the presentation
\begin{equation}\label{eq:Dic24pres}
    Dic_{24} =\langle r,s \big{|} r^{12} = s^4 = e,\, s^2=r^6, \,s^{-1} r s = r^{-1} \rangle\,.
\end{equation}
The strategy is precisely the same as in the previous Section: we start by setting up the AHSS to compute the bordism.
The integral homology of $Dic_{4n}$ with $n$ even is given by \cite{brown1982cohomology, groupprops_dicyclic}
\begin{equation}\label{eq:integral_homology_Dic4n_even}
    H_p(Dic_{4n}; \Z) = 
    \begin{cases}
        \Z \quad &p = 0,\\
        \Z_2\oplus \Z_2 \quad &p = 1,\\
        \Z_{4n} \quad &p = 3,\\
        0 \quad &p = 2k\,, \quad k\neq 0\,.
    \end{cases}
\end{equation}
Thus we can read the integral homology of $Dic_{24}$ as the $n=6$ case above:
\begin{equation}
    H_1(Dic_{24})\simeq \Z_2\oplus \Z_2, \quad H_2(Dic_{24})\simeq0, \quad H_3(Dic_{24})\simeq \Z_{24} \simeq \Z_3 \oplus \Z_8\,.
\end{equation}
The second page of the Atiyah-Hirzebruch sequence for the fibration $pt\rightarrow Dic_{24} \rightarrow Dic_{24}$ reads:

\begin{equation}\label{eq:AHSSdic24}
    E^2_{p,q} : \quad
    \begin{array}{c|cccc}
        3 & \quad & & & \\
        2 & \quad\Z_2 & \quad\Z_2\oplus\Z_2 & \quad\Z_2\oplus\Z_2 & \quad\Z_2 \\
        1 & \quad\Z_2 & \quad\Z_2\oplus\Z_2 & \quad\Z_2\oplus\Z_2 & \quad\Z_2 \\
        0 & \quad\mathbb{Z} & \quad\Z_2\oplus\Z_2 & \quad0 & \quad\Z_8 \oplus \Z_3 \\
        \hline
         q/p & 0 & 1 & 2 & 3
    \end{array}
\end{equation}
We can see that at prime 3 the only non-trivial contribution is $E^2_{3,0}\simeq \Z_3$ and that there are no possible non-trivial differentials acting on it. Thus $\Omega_3^{Spin}(BDic_{24}) \simeq \Z_3$ at prime 3. At prime 2 we again limit ourselves to observe that the maximum order of $\Omega_3^{Spin}(BDic_{24})\big{|}_{p=2}$ is 128.

Completely analogously to the previous Section, we use the structure of $Dic_{24}$ as a semidirect product
\begin{equation}
    Dic_{24} \simeq \Z_3 \rtimes Q_8\,,
\end{equation}
to write the diagram 
\begin{equation}
\begin{tikzcd}
    0 \arrow[r] & \mathbb{Z}_3 \arrow[r] & Dic_{24} \arrow[r, "\pi"] & Q_8 \arrow[r] & 0 \,,\\
        & & & Q_8 \arrow[u, equal] \arrow[ul, "i", hook] & 
\end{tikzcd}
\end{equation}
we apply the bordism functor to the triangle subdiagram and again conclude the injectivity of the map $i_*$. In other words, this means that $\Omega_3^\Spin(BQ_8)$ injects into $\Omega_3^{Spin}(BDic_{24})$ and, since the order bound at prime 2 is saturated, we can conclude that $i_*$ is an isomorphism at prime two and the full group reads
\begin{equation}\label{eq:bordismDIC24}
    \Omega_3^\Spin (BDic_{24}) \simeq \Omega_3^\Spin (BQ_8) \oplus \Omega_3^\Spin (B\Z_3)\,\simeq \Z_{24} \oplus \Z_4 \oplus \Z_4\,.
\end{equation}
The injection immediately tells us that we can take the generators of the $\Z_4$ factors to be lens spaces, just as we did for $Q_8$ in \hyperref[sec:Q8_anomaly]{Section 4}. The last generator can be constructed as a spherical space form. For this we need the representation theory of $Dic_{24}$.

There are 9 irreducible representations. Four of them are 1D, and, neglecting the trivial, read:
\begin{equation}\label{eq:Dic24_reps_1D}
    \rho_{11}:
    \begin{cases}
        \rho_{11}(r) = 1 \\
        \rho_{11}(s) = -1
    \end{cases}\,,
    \qquad \quad
    \rho_{12}:
    \begin{cases}
        \rho_{12}(r) = -1 \\
        \rho_{12}(s) = 1
    \end{cases}\,,
    \qquad\quad
    \rho_{13}:
    \begin{cases}
        \rho_{13}(r) = -1 \\
        \rho_{13}(s) = -1
    \end{cases}\,.
\end{equation}
The other five representations are all two-dimensional, three are unfaithful
\begin{equation}
    \rho_{21}(r) = \left(
        \begin{array}{cc}
         i  & 0 \\
         0 & -i \\
        \end{array}
    \right)\,, \qquad \quad
    \rho_{21}(s) = \left(
        \begin{array}{cc}
         0  & -1 \\
         1 & 0 \\
        \end{array}
    \right)\,;
\end{equation}  
\begin{equation}
    \rho_{22}(r) = \left(
        \begin{array}{cc}
         \phi  & 0 \\
         0 & \phi^{2} \\
        \end{array}
    \right)\,, \qquad \quad
    \rho_{22}(s) = \left(
        \begin{array}{cc}
         0  & 1 \\
         1 & 0 \\
        \end{array}
    \right)\,;
\end{equation} 
\begin{equation}
    \rho_{23}(r) = \left(
        \begin{array}{cc}
         -\phi  & 0 \\
         0 & -\phi^{2} \\
        \end{array}
    \right)\,, \qquad \quad
    \rho_{23}(s) = \left(
        \begin{array}{cc}
         0  & 1 \\
         1 & 0 \\
        \end{array}
    \right)\,.
\end{equation} 
The two remaining representations are faithful, and, setting $\chi=e^{\frac{2 \pi i}{12}}$ to be the twelfth root of unity, they can be written as:
\begin{equation}\label{eq:Dic24_rep_24}
    \rho_{24}(r) = \left(
        \begin{array}{cc}
         \chi  & 0 \\
         0 & \chi^{-1} \\
        \end{array}
    \right)\,, \qquad \quad
    \rho_{24}(s) = \left(
        \begin{array}{cc}
         0  & -1 \\
         1 & 0 \\
        \end{array}
    \right)\,;
\end{equation} 
\begin{equation}
    \rho_{25}(r) = \left(
        \begin{array}{cc}
         -\chi  & 0 \\
         0 & -\chi^{-1} \\
        \end{array}
    \right)\,, \qquad \quad
    \rho_{25}(s) = \left(
        \begin{array}{cc}
         0  & -1 \\
         1 & 0 \\
        \end{array}
    \right)\,.
\end{equation}
The generator of the $\Z_{24}$ summand in \eqref{eq:bordismDIC24} can be taken to be the spherical space form associated to either one of the two faithful representations, which embed in $SU(2)$. We can choose the first one in \eqref{eq:Dic24_rep_24} and build
\begin{equation}
    X_{Dic_{24}} = S^3/\rho_{24}(Dic_{24}),.
\end{equation}
The $\eta$-invariant of the same representation can be computed using \eqref{eq:etaInvariants} and evaluates to
\begin{equation}
    \tilde{\eta}(\rho_{24},X_{Dic_{24}}) = \frac{1}{24},
\end{equation}
which exactly generates the $\Z_{24}$ subgroup of $\Omega^\Spin_3(BDic_{24})$. Furthermore, we compute the $\eta$-invariants of the other representations on this generator and we get
\begin{equation}
    \tilde{\eta}(\rho_{11},X_{\rho_{24}}) = \frac{1}{2},\quad \tilde{\eta}(\rho_{12},X_{\rho_{24}}) = 0,\quad \tilde{\eta}(\rho_{13},X_{\rho_{24}}) = 0\,,
\end{equation}
for the 1D representations and
\begin{equation}
    \tilde{\eta}(\rho_{21},X_{\rho_{24}}) = \frac{3}{8},\quad \tilde{\eta}(\rho_{22},X_{\rho_{24}}) = \frac{1}{6},\quad \tilde{\eta}(\rho_{23},X_{\rho_{24}}) = \frac{2}{3}\quad \tilde{\eta}(\rho_{25},X_{\rho_{24}}) = \frac{1}{24}\,,
\end{equation}
for the 2D ones. The $\Z_4\oplus\Z_4$ factor in \eqref{eq:bordismDIC24} is generated by any two of the spaces in \eqref{eq:Q8BORD_Z4_generators}, where by $i,j,k$ we implicitly mean the images of $i,j,k$ under $i:Q_8\rightarrow Dic_{24}$.\footnote{The map in terms of the generators in \eqref{eq:Dic24pres} is given by $i\mapsto r^3, j\mapsto s, k \mapsto r^3 s $.}
The $\eta$-invariants on these generators are
\begin{equation}
    \eta(\rho_g,X_h) = \frac{1}{2}(1-\delta_{g,h}) \quad \text{for}\,\, g \in \{i,j,k\} \,\,\text{and}\,\,h\in \{i,j\}\,,
\end{equation}
for the 1D representations and
\begin{equation}
\begin{split}
    &\tilde{\eta}(\rho_{21},X_{i})= \tilde{\eta}(\rho_{21},X_{j}) = \frac{3}{4},\\
    &\tilde{\eta}(\rho_{22},X_{g}) = \tilde{\eta}(\rho_{23},X_{g}) = \frac{1}{2}(1-\delta_{g,i}) \quad \text{for}\,\, g \in \{i,j\},\\
    &\tilde{\eta}(\rho_{24},X_{i})= \tilde{\eta}(\rho_{25},X_{i}) = \tilde{\eta}(\rho_{24},X_{j})= \tilde{\eta}(\rho_{25},X_{j}) = -\frac{3}{4}\,,
\end{split}
\end{equation}
for the 2D ones.

\subsection{\texorpdfstring{$SL(2,3)$}{SL23}}

To compute anomalies of $SL(2,3)$ it is convenient to use its presentation as a semidirect product
\begin{equation}
    SL(2,3) \simeq Q_8 \rtimes \Z_3\,,
\end{equation}
where the $\Z_3$ factor acts by permuting the three generators $i,j,k$ of $Q_8$. We call this  generator $t$.\\
A presentation can then be written as follows
\begin{equation}
    SL(2,3)=\langle i,j,t \big{|} i^4 = j^4 = (ij)^4 = t^3 = 1, i^2 = j^2 = (ij)^2, t^{-1}it = j, t^{-1}jt= ij \rangle\,.
\end{equation}

There are 7 irreducible representations. Apart from the trivial one, there are two 1D ones we call \textbf{1$_\textbf{A}$} and \textbf{1$_\textbf{B}$} defined by assigning
\begin{equation}
    \rho_\textbf{1A}(t) = e^{2 \pi i/3}  \quad \text{ or } \quad \rho_\textbf{1B}(t) = e^{-2 \pi i/3}\,,
\end{equation}
and representing everything else trivially.

The three inequivalent 2D representations are faithful, and all use \eqref{eq:taurep} for the quaternionic generators, supplemented by
\begin{equation}
    \rho_{2}(t) = \frac{1}{2} \begin{pmatrix} -1+i & 1+i \\ -1+i & -1-i \end{pmatrix}\,,
\end{equation}
for the first one. The others are obtained as
\begin{equation}
    \rho_{\textbf{2A}} = \rho_2 \otimes \rho_\textbf{A}, \qquad\rho_{\textbf{2B}} = \rho_2 \otimes \rho_\textbf{B}\,.
\end{equation}
Finally, there is a 3D non-faithful representation $\rho_\textbf{3}$ generated by
\begin{equation}
    \rho_\textbf{3}(i) = \left(
\begin{array}{ccc}
 1 & 0 & 0 \\
 0 & -1 & 0 \\
 0 & 0 & -1 \\
\end{array}
\right), \qquad
\rho_\textbf{3}(j) = \left(
\begin{array}{ccc}
 -1 & 0 & 0 \\
 0 & 1 & 0 \\
 0 & 0 & -1 \\
\end{array}
\right), \qquad
\rho_\textbf{3}(t) = \left(
\begin{array}{ccc}
 0 & 1 & 0 \\
 0 & 0 & 1 \\
 1 & 0 & 0 \\
\end{array}
\right)\,.
\end{equation}

The bordism group we need is \cite{Davighi:2022icj}:
\begin{equation}
    \Omega_3^\Spin(BSL(2,3))\simeq \Z_8 \oplus \Z_3 \simeq \Z_{24}\,.
\end{equation}
The single generator can be taken to be the spherical space
\begin{equation}
    X_{SL(2,3)}:= S^3/\rho_2\,.
\end{equation}

On it, we can use \eqref{eq:etaInvariants} to compute the $\eta$-invariants of all representations on this generator. What we find is
\begin{equation}
    \begin{aligned}
        \widetilde{\eta}(\rho_\textbf{1A},X_{SL(2,3)}) = \widetilde{\eta}(\rho_\textbf{1B},X_{SL(2,3)}) &= \frac{2}{3} \,,\\ \widetilde{\eta}(\rho_\textbf{2},X_{SL(2,3)})=
    \widetilde{\eta}(\rho_\textbf{2A},X_{SL(2,3)}) = \widetilde{\eta}(\rho_\textbf{2B},X_{SL(2,3)}) &= -\frac{1}{24}\,, \\
    \widetilde{\eta}(\rho_\textbf{3},X_{SL(2,3)}) &= -\frac{1}{6}\,.
    \end{aligned}
\end{equation}
Notice that $1/24$ factor for the faithful representations signals that this is indeed a generator of the bordism group.

\subsection{\texorpdfstring{$G^7_{36}=(\Z_3\times \Z_3) \rtimes \Z_4$}{G736}}

The last group in our list which is not a direct product of groups we already examined has order 36, and we call it $G^7_{36}$ following the classification on the Small Groups Library \cite{SmallGroups}. 
We can take the following presentation
\begin{equation}
    G^7_{36} = \langle a,b,x \big{|} a^3 = b^3 = x^4 = 1, ab = ba, xax^{-1} = a^{-1}, xbx^{-1}= b^{-1} \rangle\,.
\end{equation}
To compute its bordism group $\Omega_3^\Spin(BG^7_{36})$, we again use the AHSS, this time that associated to the fibration $B\Z_3 \times B\Z_3\rightarrow BG^7_{36} \rightarrow B\Z_4$ obtained from the semidirect product structure
\begin{equation}\label{eq:z3z3z4sequence}
    0 \longrightarrow \Z_3 \times \Z_3 \longrightarrow G \longrightarrow \Z_4 \longrightarrow 0\,.
\end{equation}
For that, we need the homology of the infinite-dimensional lens space $B\Z_4$:
\begin{equation}
    H_p(B\Z_4) = \begin{cases}
        \Z &p=0 \\
        \Z_4 &p \,\text{ odd}\\
        0 &p>0, \text{ even}
    \end{cases}\,;
\end{equation}
and the bordism groups $\Omega_q^\Spin(B\Z_3 \times B \Z_3)$, which read (see e.g.~\cite{Larotonda:2026hxy} for the $d=3$ case, lower degrees are also straightforward applications of the AHSS)
\renewcommand{\arraystretch}{1.2}
\begin{equation}
    \begin{tabular}{c|c|c|c|c}
    $d$ & 0 & 1 & 2 & 3 \\
    \hline
    $\Omega_d^\text{Spin}(B\mathbb{Z}_3 \times B\mathbb{Z}_3)$ & $\Z$ & $\Z_3^{\oplus2} \oplus \Z_2$ & $\Z_3 \oplus \Z_2$ & $\Z_3^{\oplus 3}$ \\
\end{tabular}
\end{equation}
Through the UCT we set up the second page
\begin{equation}
    E_{p,q}^2 = H_p\left(B\Z_4;\Omega_q^\Spin(B\Z_3 \times B\Z_3)\right)\,,
\end{equation}
which reads
\begin{equation}\label{eq:E2z3z3z4}
    E^2_{p,q} : \quad
    \begin{array}{c|cccc}
        3 & \quad \Z_4 & \quad \Z_2& \quad \Z_2 & 0 \\
        2 & \quad 0 & \quad\Z_2 & \quad \Z_2 & 0 \\
        1 & \quad\Z_4 & \quad\Z_2 & \quad\Z_2 &  0 \\
        0 & \quad\mathbb{Z} & \quad\Z_3^{\oplus2}\oplus\Z_2 & \quad\Z_3 \oplus \Z_2 & \quad \Z_3^{\oplus3} \\
        \hline
         p/q & 0 & 1 & 2 & 3
    \end{array}
\end{equation}
From here we proceed as in the previous examples: the answer at prime 3 is read off straightforwardly as
\begin{equation}
    \Omega_3^{Spin}(BG^7_{36})\big{|}_{p=3} \simeq \Z_3 \oplus \Z_3 \oplus \Z_3 \,,
\end{equation}
while at prime 2 we limit ourselves to bound the order of the group at 16. Then as usual we supplement the sequence \eqref{eq:z3z3z4sequence} with the inclusion of the $\Z_4$ subgroup to give  
\begin{equation}
\begin{tikzcd}
    0 \arrow[r] & \Z_3 \times \Z_3 \arrow[r] & G^7_{36} \arrow[r, "\pi"] & \Z_4 \arrow[r] & 0 \,.\\
        & & & \Z_4 \arrow[u, equal] \arrow[ul, "i", hook] & 
\end{tikzcd}
\end{equation}
Applying the bordism functor to the triangle subdiagram gives again an injection
\begin{equation}\label{eq:z4inz3z3z4}
    i_*: \Omega_3^\Spin(B\Z_4) \hookrightarrow \Omega_3^\Spin(BG_{36}^7) \,.
\end{equation}
Since this saturates the bound we derived from the AHSS, we conclude
\begin{equation}
    \Omega_3^\Spin(BG_{36}^7) \simeq \Z_3^{\oplus3} \oplus \Z_8 \oplus \Z_2 \,,
\end{equation}
where the generators of the last two factors are inherited from \eqref{eq:z4inz3z3z4} and are therefore 3D lens spaces with either choice of spin structure
\begin{equation}
    L_4^3 = S^3/\Z_4, \widetilde{L}_4^3\,.
\end{equation}
As for the three generators of the $\Z_3^{\oplus 3}$ summand, they are given by lens spaces
\begin{equation}
    L_3^3 = S^3/\Z_3\,,
\end{equation}
with holonomies valued in the cyclic subgroups of the normal subgroup $\Z_3\times \Z_3 \subset G$.

In fact the map
\begin{equation} \label{eq:z3z3inj}
    i'_*: \Omega_3^\Spin(B\Z_3\times B\Z_3)\longrightarrow\Omega_3^\Spin(BG_{36}^7)
\end{equation}
is a surjection when localized at prime 3. This is not necessarily the case in general for semidirect products of the form \eqref{eq:z3z3z4sequence}.

The proof goes through a so-called transfer map, defined as in \cite{BECKER19751}. This is a map ``going the other way" with respect to \eqref{eq:z3z3inj}:
\begin{equation}
    t: \Omega_3^\Spin(BG^7_{36})\rightarrow\Omega_3^\Spin(B\Z_3 \times B\Z_3)
\end{equation}
with the property that its composition with $i'_*$ is the identity on $\Omega_3^\Spin(BG_{36}^7)$\footnote{
Property \eqref{eq:transferProp} is standard as a map in singular cohomology $H^*(BG)$ \cite{MR1867354}. For bordism, the argument goes as follows. Reference \cite{BECKER19751} defines, for a fiber bundle $F\hookrightarrow E\xrightarrow[]{\pi} B$, a stable map of spectra $\tau:\Sigma^\infty B_+\rightarrow \Sigma^\infty_+E$ from the basis to the total space such that the composition $\pi_* \circ\tau_*$ acts by multiplication by $\chi(F)$ on all generalized cohomology theories, such as spin bordism, which is what we need. Here $\chi(F)$ is the Euler number of the fiber $F$. The sequence \eqref{eq:z3z3z4sequence} gives $B\Z_3 \times B\Z_3$ the structure of a fiber bundle over $BG_{36}^7$ with fiber the coset $G_{36}^7/(\Z_3\times \Z_3)\simeq \Z_4$. The composite map is then multiplication by $\chi(\Z_4) = 4$, and since we know the bordism at prime 3 is given by $\Z_3^{\oplus3}$, in this group multiplying by 4 is the identity. }:
\begin{equation}\label{eq:transferProp}
    i'_* t = id: \Omega_3^\Spin(BG_{36}^7)\rightarrow \Omega_3^\Spin(BG_{36}^7)\,,
\end{equation}
which through arguments similar to those above implies $i'_*$ is surjective, and thus all order 3 generators of $\Omega_3^\Spin(BG_{36}^7)$ are directly obtained by using $\Z_3\times\Z_3$ bundles.

To summarize, a full set of bordism generators  is given by
\begin{equation}
     \Omega_3^\Spin(BG) = \langle  L_4^3, \widetilde{L}_4^3, L_3^3[a],L_3^3[b],L_3^3[ab]  \rangle \,,
\end{equation}
where by the notation $L_3^3[a]$ we mean that we have turned on an order 3 holonomy for the element $a$, and analogously for $L_3^3[b],L_3^3[ab]$.

Since all generators are induced by abelian subgroups, we conclude that level matching is sufficient to guarantee the absence of anomalies.

\section{Full classification of \texorpdfstring{$T^5\times S^1$}{T5xS1} orbifolds}\label{sec:5Dresults}

In this work we solve the classification problem of point groups with a simplifying assumption, both conceptually and computationally. A subclass of $T^6$ orbifolds is given by actions that factorize as $T^5 \times S^1$, and act purely as geometric shifts on the $S^1$ factors. The simplification in the search is twofold: first of all, we are looking into representations of symmetry groups of 5D lattices, which are far fewer than their 6D counterparts; secondly, most twisted sectors can be guaranteed to be devoid of massless states as follows.

Since there is no asymmetric action on the $S^1$, the radius is not going to be fixed, and in particular it can be taken to be large. It is well known that in geometric compactifications massless twisted states reside at points which are fixed by some group element \cite{Dixon:1985jw}. By taking the circle radius to be sufficiently bigger than the string scale $\sqrt{\alpha'}$, a simple calculation \cite{Baykara:2023plc} shows that any sector twisted by an element with a shift on the circle has no massless states.

We therefore have to figure out what elements have no associated shifts, in each of our orbifolds. Abstractly, this can be done as follows. Since shifts on the circle do not mix with the internal $T^6$, the assignement of a shift to each group element defines a homomorphism
\begin{equation}
    \chi : G \rightarrow U(1)
\end{equation}
and we can describe the elements which do not have a shift as its kernel, which is a normal subgroup
\begin{equation}
    N = \text{ker}(\chi)\,.
\end{equation}

All shifts necessarily commute with each other, so this map factors through the abelianization $G_{ab}= G/[G,G]$. This means that certainly $[G,G] \subset N$. To see if there are any additional elements in $N$ we have to study the kernel of the induced map $\widetilde{\chi}:G_{ab}\rightarrow U(1)$. In particular, since any finite subgroup of $U(1)$ is cyclic, if $G_{ab}$ is not cyclic itself, then ker$(\widetilde{\chi})$ cannot be trivial. In the cases where it is not trivial, we build $N$ as the subgroup generated by $[G,G]$ and a representative for every class in ker$(\widetilde{\chi})$.

Only sectors twisted by elements of $N$ can give rise to massless gravitini, signalling that the orbifold is secretly supersymmetric. We know from the general analysis of \hyperref[sec:as_orbifolds]{Section 2} that gravitini can only arise in such a sector if the associated point group element acts trivially either on the left or on the right. We will examine cases in which this happens individually.

The point groups are also greatly restricted compared to their 6D counterparts. We implement this in our search by adding an extra condition to \textbf{I.}-\textbf{VI.} above:
\begin{itemize}
    \item[] \textbf{VII
    .} The induced bosonic representations have a trivial subrepresentation both on the left and on the right.
\end{itemize}
In other words, this means that the induced representation really lands in an $SO(5)$ subrepresentation rather than the full $SO(6)$. By knowing that our fermionic representations should project down to a crystallographic symmetry of a 5D lattice, together with the knowledge that the extension $0\rightarrow \Z_2 \rightarrow G_F \rightarrow G_B \rightarrow0$ should be trivial, as argued in \hyperref[sec:rep_th]{Section 3} and \hyperref[app:ext]{Appendix B}, we can strongly restrict the list of groups we should even consider as candidates in the first place. To get an exhaustive list we proceed as follows.

Symmetries of 5D lattices are finite proper subgroups of $GL(5;\Z)$. There are 17 maximal subgroups, corresponding to symmetries of highly symmetric crystals, see e.g.~\cite{Veysseyre:js0121}. Definitions of the concepts used in this classification, and in CARAT at large, are reviewed in \hyperref[app:cyclotomic]{Appendix A}. By maximal we mean that their automorphism group can not be embedded, up to conjugation in $GL(5;\Z)$, into the automorphism group of some other lattice. To find them, we use CARAT to get a full list of Bravais groups of 5D crystal systems, and then use GAP to verify which ones are maximal. The results can be found in \hyperref[tab:crystals]{Table 2}. Most of these lattices can be written\footnote{Up to overall scaling, which is irrelevant for us since asymmetric actions are symmetries only at the self-dual radius, which then fixes the scaling.} as root or weight lattices of a semi-simple Lie algebra. Their symmetry groups are therefore nothing but the Weyl group of the associated algebra, possibly extended by some outer automorphism.

\begin{table}[!ht]
    \centering \label{tab:crystals}
    \begin{tabular}{|c|c|c|c|c|}
    \hline
    \textbf{\#} & \textbf{Crystal Family} & \textbf{Description} & \textbf{Automorphism group} & \textbf{Order}  \\ \hline
    \textbf{1} & \multirow{3}{*}{5-1} & $\Lambda_R(B_5) \simeq \sqrt{2} \,\Z^5$ & $W(B_5) \simeq \mathbb{Z}_2 \wr S_5$ & 3840 \\ 
    \textbf{2} & & $\Lambda_R(D_5)$ & $W(B_5)$ & 3840 \\ 
    \textbf{3} & & $\Lambda_W(D_5)$ & $W(B_5)$ & 3840 \\ \hline
    \textbf{4} & 4-1;1& $\Lambda_R(F_4 \oplus A_1)$& $W(F_4) \times \Z_2$ & 2304\\ \hline
    \textbf{5} & \multirow{4}{*}{5-2} & $\Lambda_R(A_5)$& $W(A_5) \times \mathbb{Z}_2 \simeq S_6 \times \mathbb{Z}_2$ & 1440 \\ 
    \textbf{6} & & $\Lambda_W(A_5)$& $W(A_5) \times \mathbb{Z}_2$ & 1440 \\ 
    \textbf{7} & & $A_5^2$& $W(A_5) \times \mathbb{Z}_2$ & 1440 \\
    \textbf{8} & & $A_5^3$& $W(A_5) \times \mathbb{Z}_2$ & 1440 \\ \hline
    \textbf{9} & \multirow{3}{*}{3;2-2}& $\Lambda_R(B_3) \oplus \Lambda_R(A_2) $& $W(B_3\oplus A_2)\simeq (\Z_2 \wr S_3) \times D_{12}$ & 576 \\
    \textbf{10} & & $\Lambda_R(D_3) \oplus \Lambda_R(A_2) $& $W(B_3\oplus A_2)$ & 576 \\ 
    \textbf{11} & & $\Lambda_W(D_3) \oplus \Lambda_R(A_2) $& $W(B_3\oplus A_2)$ & 576 \\ \hline
    \textbf{12} & \multirow{2}{*}{4-2;1}& $\Lambda_R(A_2\oplus A_2 \oplus A_1) $& $(W(A_2) \times W(A_2)) \rtimes \mathbb{Z}_2$ & 576 \\ 
    \textbf{13} & & $\Gamma^4(A_2)$& $(S_3 \wr S_2)\times \Z_2$ & 288 \\ \hline
    \textbf{14} & \multirow{2}{*}{4-3;1}& $\Lambda_R(A_4)\oplus \Lambda_R(A_1)$& $W(A_4) \times \mathbb{Z}_2$ & 480 \\ 
    \textbf{15} & & $\Lambda_W(A_4)\oplus \Lambda_R(A_1)$& $W(A_4) \times \mathbb{Z}_2$ & 480 \\ \hline
    \textbf{16} & \multirow{2}{*}{3;2-1}& $\Lambda_R(D_3)\oplus \Lambda_R(B_2)$& $W(D_3 \oplus B_2) \simeq (\Z_2 \wr S_3) \times D_8$  & 384 \\ 
    \textbf{17} & & $\Lambda_W(D_3)\oplus \Lambda_R(B_2)$& $W(D_3 \oplus B_2)$ & 384 \\ \hline
    \end{tabular}
    \caption{ The 5D lattices whose automorphism group is maximal, together with their CARAT crystal family and their automorphism group. Most lattices can be expressed as root or weight lattices of semi-simple Lie algebras, up to overall scaling which for us is fixed at the self-dual radius. The lattices $A_5^2$ and $A_5^3$ are obtained by adjoining to $\Lambda_R(A_5)$ representative elements of the two proper subgroups of $\Lambda_W(A_5)/\Lambda_R(A_5) \simeq \Z_2 \oplus \Z_3$, of order 2 and 3 respectively. $\Gamma^4(A_2)$ is obtained by gluing, as in \eqref{eq:extraSymm}, from $\Lambda_R(A_2)$ and equipping it with a positive definite quadratic form.}
\end{table}

With GAP we can directly compute and list all proper subgroups of these maximal groups. Since only the two maximal subgroups have order bigger than $2000$, and they both are lower than $4000$, this ensures that every proper subgroup is in the GAP Small Groups Library, and we can assign them their ID from there. This ID is what we will use to generate and label irreducible representations.

Using GAP and this list, we can build all possible fermionic representations which satisfy properties \textbf{I.-IV.} and \textbf{VI.} above. The property which needs some more care is \textbf{V.}. To ensure it when we project the fermionic representation down to bosons, we have built through the CARAT package of GAP a full list of the representations that appear in the lattice symmetry groups in \hyperref[tab:crystals]{Table 2}, in terms of their decomposition into irreducible representations of the group that they represent faithfully. Then once we have found a fermionic representation, we compute the characters of the induced bosonic representation, and compare it with these lists. If both the left-moving and right-moving bosonic representations appear in a list associated to the same Bravais group, this means that they both are symmetries of the same lattice, and we can find a basis to write the transformation as integral. We only do this last step explicitly in the most straightforward cases, as it is in general non-trivial to do algorithmically.

For those models for which we cannot find an integral representation, we limit ourselves to presenting the decomposition into irreducible complex representations. The search algorithm we just described guarantees that for each lattice we list, such an integral representation does exist.

Curiously, the anomaly cancellation conditions, calculated with the tools of \hyperref[sec:anomalies]{Section 4}, are satisfied by all solutions to conditions \textbf{I.}-\textbf{VI.}.
This suggests a strong correlation between the property of vanishing one-loop partition function and the cancellation of anomalies even in the non-abelian case, but we do not have a general proof of this fact.

At least in the abelian case, modular invariance at one loop is enough to guarantee it at all loops \cite{Freed:1987qk}, so heuristically setting the partition function to zero automatically makes it well-defined at one loop, and therefore at all loops. It would be interesting to refine this intuition into a precise argument for any orbifold group $G$.

Before describing examples in detail, we summarize the results of the classification here.
We find 17 orbifolds with vanishing partition function, listed in \hyperref[tab:results]{Table 1}; for all but the $G=S_3$ model, no shifts beyond the geometric one on the base $S^1$ are required. 

For $G=S_3$, additional shifts are needed to remove massless twisted gravitini, but they can be chosen without modifying the group structure. By this we mean that the space group $G$ remains isomorphic to the point group $P_G$, thus no extra sectors are added in the orbifold sum \eqref{eq:orbifold_partition_function} and the analysis up to this point is enough to guarantee the vanishing of the vacuum energy at one loop.
In this case, in which the choice of shift is important, we do not classify all of the possibilities, and we limit ourselves to one choice of lattice and one choice of shift on that lattice which is consistent.

\begin{table}[t]
\renewcommand{\arraystretch}{1.25}\centering
\begin{tabular}{|
c|
@{}>{\,}c<{\,}@{}|
@{}>{\,$}c<{$\,}@{}|
@{}>{\,$}c<{$\,}@{}|
@{}>{\,$}c<{$\,}@{}|
>{$}c<{$}|} \hline
$d$ &\text{Id(G)} & {G} & G_F & G_B & \text{\#} \\ \hline
\multirow{7}{*}{4} & \multirow{2}{*}{$[16,12]$} & \text{\multirow{2}{*}{$\Z_2 \times  Q_8$}} & {\Z_2^2}\oplus(\Z_2 \times  Q_8) & {\Z_2}\oplus Q_8 &3\\
& && {\Z_2^2}\oplus Q_8 & {\Z_2}\oplus Q_8 &3\\\cline{2-6}
& \multirow{2}{*}{$[24,7$]} & \text{\multirow{2}{*}{$\Z_2 \times  Dic_{12}$}} & {\Z_2^2}\oplus(\Z_2 \times  Dic_{12}) & {\Z_2}\oplus Dic_{12}&1 \\
& && {\Z_2^2}\oplus Dic_{12} & {\Z_2}\oplus Dic_{12}&1 \\\cline{2-6}
& $[32,47]$ & {\Z_2^2 \times  Q_8} & {\Z_2^2}\oplus(\Z_2 \times  Q_8) & {\Z_2}\oplus Q_8& 1 \\\cline{2-6}
& $[48,42]$ & {\Z_2^2 \times  Dic_{12}} & {\Z_2^2}\oplus(\Z_2 \times Dic_{12}) & {\Z_2}\oplus Dic_{12}& 1 \\\cline{2-6}
& $[96,198]$ & {\Z_2^2 \times SL(2,3)} & {\Z_2^2}\oplus(\Z_2 \times SL(2,3)) & {\Z_2}\oplus{SL(2,3)}& 2 \\
 \hline
\multirow{18}{*}{5} &\multirow{2}{*}{$[12,4]$} & \text{\multirow{2}{*}{$D_{12}$}} & D_{12} \oplus \Z_2 & S_3\oplus 1 &1\\ 
& && D_{12}\oplus {\Z_2^2} & { S_3 }\oplus{\Z_2} &2 \\\cline{2-6}
&  $[24,5]$ & {\Z_4 \times  S_3 } &  D_{12}\oplus (\Z_4 \times \Z_2) & S_3\oplus \Z_4 &1\\\cline{2-6}
&  \multirow{2}{*}{$[24,7$]} & \text{\multirow{2}{*}{$\Z_2 \times  Dic_{12}$}} & D_{12}\oplus {\Z_4} & S_3\oplus {\Z_4} &1\\
&  &&  D_{12}\oplus Dic_{12} &  S_3\oplus Dic_{12}&1\\\cline{2-6}
& $[24,14]$ & 
  {\Z_2 \times D_{12}} & D_{12}\oplus {\Z_2^2}  & {S_3}\oplus{ \Z_2 }& 1 \\\cline{2-6}
&  \multirow{3}{*}{$[36,12]$} & \text{\multirow{3}{*}{$ \Z_6 \times  S_3$ }} & D_{12}\oplus\Z_6 & S_3\oplus\Z_3& 1 \\
&  & & D_{12}\oplus\Z_6 & S_3 \oplus \Z_6& 1 \\
&  & &  D_{12}\oplus{ (\Z_6 \times \Z_2)} & { S_3 }\oplus{ \Z_6}& 2 \\\cline{2-6}
&  $[48,34]$ & {\Z_2 \times Dic_{24}} &  D_{12}\oplus Q_8 &  S_3 \oplus Q_8& 1 \\\cline{2-6}
&  $[48,35]$ & {
  \Z_4 \times D_{12}} &  D_{12}\oplus (\Z_4 \times \Z_2) & S_3\oplus{\Z_4}& 1 \\\cline{2-6}
&  $[48,40]$ & { S_3 \times  Q_8 } &  D_{12}\oplus(\Z_2 \times  Q_8) & { S_3 }\oplus Q_8& 3 \\\cline{2-6}
&  $[72,20]$ & { S_3 \times Dic_{12} } &  D_{12}\oplus(\Z_2 \times Dic_{12}) & { S_3 }\oplus Dic_{12}& 1 \\\cline{2-6}
&  $[72,34]$ & \Z_2 \times G^7_{36} &  D_{12}\oplus Dic_{12} &  S_3 \oplus Dic_{12} & 1 \\\cline{2-6}
&  $[72,48]$ & {
  \Z_6 \times D_{12}} &  D_{12}\oplus(\Z_6 \times \Z_2) & { S_3 }\oplus{ \Z_6}& 1 \\\cline{2-6}
&  $[96,212]$ & {D_{12} \times  Q_8 } &  D_{12}\oplus(\Z_2 \times  Q_8) & { S_3 }\oplus Q_8& 1 \\\cline{2-6}
&  $[144,146]$ & {D_{12} \times  Dic_{12}} &  D_{12}\oplus(\Z_2 \times Dic_{12}) & { S_3 }\oplus Dic_{12}& 1 \\\cline{2-6}
&  $[288,922]$ & {D_{12}  \times SL(2,3)} &  D_{12}\oplus(\Z_2 \times SL(2,3)) & { S_3 }\oplus{SL(2,3)} & 2 \\ \hline
 \end{tabular}\caption{Orbifolds satisfying conditions \textbf{I.-V.} but not \textbf{VI.} Without shifts, the partition function is not zero, but the space group may in principle be enhanced to a solution with $V_{1-\text{loop}}=0$. The orbifolds are grouped based on the dimension $d$ of the internal torus $T^d$ on which they act faithfully. The last column indicates the number of inequivalent orbifolds for each $G$.}

\label{tab:listT5S1_ItoV}
\end{table}

We also found a larger class of models which satisfy conditions \textbf{I.}-\textbf{V.} but not condition \textbf{VI.}. They are summarized in \hyperref[tab:listT5S1_ItoV]{Table 3}. This means that some non-trivial commuting pairs do not preserve any common supercharge, despite all elements of the group preserving some on their own. Without any shifts, the vacuum energy will not be zero in these cases. Let us stress that this is not necessarily a killer, as it is in fact precisely the situation in which the abelian models of \cite{Larotonda:2026hxy} find themselves. What one could do to still have $V_{1-\text{loop}}=0$ in such an orbifold is add shifts such that the space group is enhanced to a larger non-abelian group $\widetilde{G}\supset G$ in which the ``problematic" commuting pairs do not commute anymore. While a priori possible, this in general introduces many more sectors in \eqref{eq:orbifold_partition_function}, and more possibilities for non-abelian anomalies to arise (see e.g.~\cite{Larotonda:2026hxy}, where for example the point group $\Z_3 \times \Z_3$ was enhanced to a non-abelian space group, $\Delta(27)$). Even if some solutions may exist, we do not expect it to be possible in general. This enhancement complicates the analysis considerably so we do not attempt it in the present work.

Before delving into the examples, let us discuss an issue that might have arisen in a more general version of our procedure. For each model, we started from a fermionic representation $\rho_F$ of the point group, found the induced bosonic representation $\rho_B$, still of the point group, and only then added shifts. We did so in a way that the space group was not enhanced at the level of bosons, which is straightforward to check. At a fermionic level, the absence of an enhancement is less trivial. In fact, while shifts locally do not act on fermions, they may induce holonomies around non-trivial cycles. In particular, this is the case if the spin structure on the parent $T^6$ is not chosen to be the fully periodic one \cite{ValeixoBento:2025yhz}. To exemplify what could go wrong, suppose one has an element $g$ of order 3 in $G_F \simeq \Z_3$, and equips it with an order 3 shift on the base circle that has antiperiodic boundary conditions. Then $g^3$ is a pure shift around the circle, which therefore acts with a $(-1)^F$ holonomy. Then $g^3\neq 1$ on fermions, and in fact we had an enhancement $G_F\rightarrow\widetilde{G}_F \simeq \Z_6$.
Since there is no action of the point group on the base $S^1$, we are allowed to always choose the periodic spin structure on it \cite{ValeixoBento:2025yhz}, and no such enhancement can occur.

To write generators, we borrow notation from \cite{Larotonda:2026hxy} and write them in ``mostly diagonal form". For example, the generator  a fully diagonal order $k$ fermionic action is going to be written as
\begin{equation}\label{eq:f}
    f = (z^{f_1},z^{f_2},z^{f_3},z^{f_4}|z^{f_5},z^{f_6},z^{f_7},z^{f_8}), \quad \text{with}\, z = e^\frac{2 \pi i}{k},
\end{equation}
where $z=e^{2 \pi i/k}$ is a $k$-th root of unity. When the action is not diagonal, the notation \eqref{eq:f} is replaced with a block diagonal form, where the blocks are fully explicit.

\subsection{\texorpdfstring{$S_3$}{S3}}\label{sec:S3}
The smallest non-abelian group, $S_3$, admits a representation respecting our properties. It is unique and acts as a $\Z_2$ on the right and an $S_3$ on the left. We can write the generators of the fermionic action as
\begin{equation}
    \begin{aligned}
        &r = \left(1,1,1,1 \big{|}\phi,\phi^{-1},\phi,\phi^{-1}\right)\,, \\
        &s = \left(-1,-1,-1,-1 \big{|} \SmallMatrix{0 & 1 \\1 & 0},\SmallMatrix{0 & 1 \\1 & 0}\right) \, ,
    \end{aligned}
\end{equation}
where $\phi = e^{2 \pi i/3}$ is a third root of unity.

The induced bosonic action is a symmetry of all lattices \textbf{1}-\textbf{17} in \hyperref[tab:crystals]{Table 2}. For concreteness we focus on a solution of the form:
\begin{equation}
    \Lambda = \Lambda_R(A_2) \oplus \Lambda_3 \oplus \Lambda_1 \,,
\end{equation}
with $\Lambda_3$ and $\Lambda_1$ arbitrary lattices of rank 3 and 1 respectively.

The bosonic action on this lattice is completely trivial on the left, only leaving $s$ acting as $(-1)^{F_L}$, while on the right it is given by
\begin{equation}
    \begin{aligned}
        r_R &= \left(\SmallMatrix{ 0 & -1 \\
        1 & -1},1,1,1,1\right)\otimes (-1)^{F_R} \, , \\
    s_R &= \left(\SmallMatrix{-1 & 1 \\
        0 & 1},-1,-1,-1,1\right) \, .
    \end{aligned}
\end{equation}

$S_3$ does not have any commuting pairs which are not of the form $(g^a,g^b)$, so we conclude that the vacuum energy at one loop is zero without any need of further checks.

This point group action has elements acting trivially on left-moving bosons (in fact all elements), so as presented, the model will have gravitini in twisted sectors, and we therefore need to add shifts.

The solution to lift their masses is precisely the same as in the $S_3 \times \Z_3$ example of \cite{Larotonda:2026hxy}, since the right-moving bosonic action is essentially the same.
We can add a geometric order 2 shift in the base circle $\Lambda_1$ for $s$, and an order 3 shift to $r$ in the $\Lambda_3$ component on the right. Note that the commutator subgroup of $S_3$ is $[S_3,S_3]\simeq \Z_3 = \langle r \rangle$, so we are not allowed to put a base circle shift on $r$ without spoiling the group structure.

The twist vector associated to $r$ is
\begin{equation}
    \vec{\theta}_R(r) = \left( \frac{4}{3},0,0 \right)\,,
\end{equation}
so we can take as a concrete example of a lattice $\Lambda_3 = \Lambda_R(A_2) \oplus \Lambda_R(A_1)$ and add a shift exactly as in \cite{Larotonda:2026hxy}:
\begin{equation}\label{eq:order3shift}
    v_2 = \frac{1}{3}\alpha_1 + \frac{2}{3} \alpha_2 \in \Lambda_W(A_2) \subset\frac{1}{3}\Lambda_R(A_2) \,,
\end{equation}
where $\alpha_i$ are the simple roots of $A_2$.
This does not spoil level matching, so it does not introduce any anomaly for the chiral bosons, and one can use the mass formulae \eqref{eq:Hvec} and \eqref{eq:Hspinor} to directly check that all the potential gravitino masses are strictly positive. In fact it suffices to check the $r$-twisted sector, since every other sector is lifted by the geometric shift on $S^1$. The shift \eqref{eq:order3shift} lifts the corresponding mass.

As far as non-abelian anomalies, as remarked above and shown in \cite{Larotonda:2026hxy}, we do not need to check anything apart from level matching, so the model is modular invariant.

\subsection{\texorpdfstring{$Q_8$}{Q8}}\label{sec:Q8}

Next up is the quaternion group $Q_8$. There are three fermionic representations satisfying our conditions. They all act as a $\Z_2\times\Z_2$ on the left and faithfully as $Q_8$ on the right movers. These are given as

\begin{equation}
    \begin{aligned}
        \rho^F_1 : \begin{cases}
            i = \left(-1,-1,1,1 \big{|}-1,-1, \SmallMatrix{
        0 & -1 \\
        1 & 0
   }\right) \\
        j = \left(1,1,-1,-1 \big{|}-1,-1,i,-i\right)
        \end{cases}\,, \\
        \rho^F_2 : \begin{cases}
            i = \left(-1,-1,-1,-1 \big{|}1,1, \SmallMatrix{
        0 & -1 \\
        1 & 0
   }\right) \\
        j = \left(1,1,-1,-1 \big{|}-1,-1,i,-i\right)
        \end{cases} \,,\\
    \rho^F_3 : \begin{cases}
            i = \left(-1,-1,1,1 \big{|}-1,-1, \SmallMatrix{
        0 & -1 \\
        1 & 0
   }\right) \\
        j = \left(-1,-1,-1,-1 \big{|}1,1,i,-i\right)
        \end{cases}\,.\\
    \end{aligned} 
\end{equation}
We only examine the first one in detail here. The same steps can be taken entirely analogously for the other two. The induced bosonic action is a symmetry of the lattices \textbf{1-4} in \hyperref[tab:crystals]{Table 2}, acting as $\Z_2$ on the left and $Q_8$ on the right. 
As an example, we choose the hypercubic lattice \textbf{1} and write:

\begin{equation}\label{eq:bosonicQ8action}
    \begin{aligned}
        \rho^B_1 : \begin{cases}
                    i = \left(1,1,-1,-1,-1,-1\big{|}1,1, \SmallMatrix{
                        0 & -1 & 0 & 0 \\
                        1 & 0 & 0 & 0 \\
                        0 & 0 & 0 & -1 \\
                        0 & 0 & 1 & 0 }\right) \otimes (-1)^{F},\\
                    j = \left(1,1,-1,-1,-1,-1\big{|}1,1,  \SmallMatrix{
                        0 & 0 & -1 & 0 \\
                        0 & 0 & 0 & 1 \\
                        1 & 0 & 0 & 0 \\
                        0 & -1 & 0 & 0 }\right)\otimes (-1)^{F_R}.
                    \end{cases}
    \end{aligned}
\end{equation}

Note that this bosonic point group acts on a $T^4$ only, leaving a full ``base" $T^2$ invariant.

We can put order 2 shifts on this $T^2$ for both generators to lift all twisted states apart from the commutator subgroup $[Q_8,Q_8]\simeq \Z_2 \simeq \langle-1\rangle$. Since it acts trivially on bosons on the left, it would seem that there are massless gravitini in the $(-1)$-twisted sector. However, we must implement the orbifold projection, and since $-1$ is central we have $\mathcal{C}(-1) = Q_8$. Then for a vector to survive on the left and be able to give a gravitino in the spectrum, we would need the action of the whole group to be trivial on bosons on the left. From \eqref{eq:bosonicQ8action} we can see this is not the case, and therefore there is no SUSY restoration from the twisted sectors.

\subsection{\texorpdfstring{$Dic_{12}$}{Dic12}}\label{sec:Dic12}

The dicyclic group of order 12 has two fermionic solutions which act crystallographically on bosons.
The first one acts as $\Z_4$ on the left, and as $S_3$ on the right. The fermionic action is generated by
\begin{equation}
    \begin{aligned}
        &r = \left(1,1,-1,-1 \big{|}\phi,\phi^{-1},\phi,\phi^{-1}\right)\,, \\
        &s = \left(-1,-1,i,-i \big{|}\SmallMatrix{0 & 1 \\1 & 0},\SmallMatrix{0 & 1 \\1 & 0}\right)\,,\end{aligned}
\end{equation}
with $\phi = e^{2 \pi i/3}$ a third root of unity. The induced bosonic action is given as
\begin{equation}
    \begin{aligned}
        &\rho_L = \mathbf{1}^{\oplus 2} \oplus \rho_{12}^{\oplus 2} \oplus \rho_{13}^{\oplus 2}  \,,\\
        &\rho_R = \mathbf{1}\oplus \rho_{11}^{\oplus 3} \oplus \rho_{22}\,,
    \end{aligned}
\end{equation}
where the representations used here are defined in \hyperref[sec:Dic_12_anomaly]{Section 4.2}.

It is a symmetry of lattices \textbf{1-4},\textbf{12},\textbf{16} and \textbf{17} in \hyperref[tab:crystals]{Table 2}.

The second solution acts as $Dic_{12}$ on the left and $S_3$ on the right. The fermionic action is generated by
\begin{equation}
    \begin{aligned}
        &r = \left(1,1,-\phi,-\phi^{-1} \big{|}\phi,\phi^{-1},\phi,\phi^{-1}\right)\,, \\
        &s = \left(-1,-1,\SmallMatrix{0 & -1 \\1 & 0}\big{|}\SmallMatrix{0 & 1 \\1 & 0},\SmallMatrix{0 & 1 \\1 & 0}\right)\,.
    \end{aligned}
\end{equation}
The induced bosonic action is a symmetry of lattice \textbf{12} in \hyperref[tab:crystals]{Table 2} and is given by
\begin{equation}
    \begin{aligned}
        &\rho_L = \mathbf{1}^{\oplus 2}\oplus \rho_{21}^{\oplus 2}\,,\\
        &\rho_R = \mathbf{1}\oplus \rho_{11}^{\oplus 3}\oplus \rho_{22}\,.
    \end{aligned}
\end{equation}

We can consistently add base circle shifts of order 4 for $s$ and of order 2 for $r$ to get rid of most twisted sectors. The only elements without shifts are those of the commutator subgroup $[Dic_{12},Dic_{12}]\simeq \Z_3$, generated by $r^2$. It acts trivially on the left, so it could a priori give gravitini in the twisted sector. However, it clearly commutes with the central $r^3$, which does not act trivially on the left, and therefore projects the candidate massless gravitino out of the spectrum.

\subsection{\texorpdfstring{$D_{12}$}{D12}}\label{sec:D12}
The dihedral group of order 12 has one fermionic representation satisfying our constraints, which was found in \cite{Larotonda:2026hxy}. It is generated at the fermionic level by
\begin{equation}
    \begin{aligned}
        &r = \left(1,1,-1,-1 \big{|}\phi,\phi^{-1},\phi,\phi^{-1}\right) \,,\\
        &s = \left(-1,-1,-1,-1 \big{|}\SmallMatrix{        0 & 1 \\1 & 0},\SmallMatrix{0 & 1 \\
        1 & 0}\right)\,,
    \end{aligned}
\end{equation}
where $\phi = e^{2 \pi i/3}$ is a third root of unity.
As it can be seen, it acts only as $\Z_2 \times \Z_2$ on the left, and faithfully as $S_3$ on the right.

The induced bosonic action is again faithful on the right while being a $\Z_2$ on the left. In general, the action on bosons is a symmetry of all lattices \textbf{1-17} in \hyperref[tab:crystals]{Table 2}.

For concreteness, we can take a lattice of the form
\begin{equation}
    \Lambda = \Lambda_4 \oplus \Lambda_R(A_2)\,,
\end{equation}
where $\Lambda_4$ is an arbitrary lattice of rank 4, and write the generators as
\begin{equation}
    \begin{aligned}
        &r = \left(1,1,-1,-1,-1,-1 \big{|}1,1,1,1,\SmallMatrix{0 & -1 \\
        1 & -1}\right) \otimes (-1)^{F_{R}}\,,\\
        &s = \left(1,1,1,1,1,1 \big{|}
        1,-1,-1,-1,\SmallMatrix{-1 & 1 \\
        0 & 1}\right)\otimes (-1)^{F_{L}}\,.
    \end{aligned}
\end{equation}

To get rid of most massless twisted sectors, we take $\Lambda_4 = \Lambda_3 \oplus \Lambda_1$ and use shifts on the base $S^1$ as usual. To preserve the group structure, we are allowed to use shifts of order 2 on both $r$ and $s$.
\\The subgroup of elements without shifts is, in this case, $N = \langle r^2, sr \rangle \simeq S_3$ .

The element $sr$ acts non-trivially both on the left and the right, so no gravitini can arise there. $r^2$ does act trivially on the left, but it clearly commutes with $r$, which projects the potential gravitino out of the spectrum.

\subsection{\texorpdfstring{$S_3 \times \Z_3$}{S3 x Z3}}\label{sec:S3Z3}

There are two fermionic representations. The first one is the one found in \cite{Larotonda:2026hxy}, the other differs just in the generator $s$. They both act as $S_3$ on the right and $\Z_6$ on the left. The fermionic actions are:
\begin{equation}
    \begin{aligned}
       &r = \left(1,1,1,1 \big{|}\phi,\phi^{-1},\phi,\phi^{-1}\right) \,,\\
        &s = \left(-1,-1,\pm 1,\pm 1\big{|}\SmallMatrix{0 & 1 \\1 & 0} \SmallMatrix{0 & 1 \\1 & 0}\right)\,,\\
        &t = \left(1,1,\phi^{-1},\phi \big{|} 1,1,1,1\right)\,,
    \end{aligned}
\end{equation}
where $\phi = e^{2 \pi i/3}$ is a third root of unity.
Both the bosonic representations act as a $S_3$ on the right, while on the left the first acts as $\Z_6$ and the second as $\Z_3$.  The first bosonic action is a symmetry of lattices \textbf{4}, \textbf{12} and \textbf{13} in \hyperref[tab:crystals]{Table 2} while the second one is a symmetry of lattices \textbf{4-13}.

For concreteness, we choose here a lattice of the form 
\begin{equation}
    \Lambda =  \Lambda_1 \oplus \Lambda_R(A_1) \oplus \Lambda_R(A_2) \oplus \Lambda_R(A_2) \,,
\end{equation}
where $\Lambda_1$ is an arbitrary lattice of rank 1.
On this lattice, we can explicitly write the bosonic actions as:
\begin{equation}\label{eq:S3xZ3_bosonic_actions}
    \begin{aligned}
        &r = \left(1,1,1,1,1,1\big{|}1,1,1,1, \SmallMatrix{0 & -1 \\
        1 & -1}\right)\otimes(-1)^{F_{R}}\,,\\
       &s = \left(1,1,\mp 1,\mp 1,\mp 1,\mp 1\big{|}1,-1,-1,-1,\SmallMatrix{-1 & 1 \\
        0 & 1}\right)\otimes(-1)^{\gamma\,  F_L}\,,\\
        & t = \left(1,1,\SmallMatrix{0 & -1 \\
        1 & -1},\SmallMatrix{0 & -1 \\
        1 & -1}\big{|}1,1,1,1,1,1\right)\,,
    \end{aligned}
\end{equation}
where $\gamma  = 0$ for the first solution, while $\gamma =1$ for the second.

The only elements without shifts are those in the commutator subgroup $[S_3\times\Z_3, S_3\times\Z_3] \simeq \Z_3 = \langle r \rangle$. So we can consistently add shifts of order 2 to $s$ and order 3 to $t$ on the base $S^1$ to get rid of massless twisted sectors. The commutator $r$ has a trivial action on the left, so a priori it could give a massless gravitino. It however commutes with the central $t$ which acts non-trivially on the left, projecting out the tentative gravitino.

\subsection{\texorpdfstring{$Dic_{24}$}{Dic24}}\label{sec:Dic24}

The Dicyclic group of order 24 has two fermionic representations which realize the mechanism we are interested in, acting crystallographically on bosons. The first one acts as a $Q_8$ on the left and $S_3$ on the right. The fermionic action is generated by:
\begin{equation}\label{eq:dic24sol}
    \begin{aligned}
        &r = \left(1,1,i,-i \big{|}\phi,\phi^{-1},\phi,\phi^{-1}\right) \,,\\
        &s = \left(-1,-1,\SmallMatrix{ 0 & -1 \\
        1 & 0} \big{|}\SmallMatrix{ 0 & 1 \\
        1 & 0},\SmallMatrix{ 0 & 1 \\
        1 & 0}\right)\,,
    \end{aligned}
\end{equation}
where $\phi = e^{2 \pi i/3}$ is a third root of unity.
This induces an analogous action on bosons that turns out to be a symmetry of lattices \textbf{1-4} in \hyperref[tab:crystals]{Table 2}. This decomposes in terms of irreducible representations of $Dic_{24}$, defined in \hyperref[sec:Dic_24_anomaly]{Section 4.3}, as
\begin{equation}
    \begin{aligned}
        & \rho_L = \mathbf{1}^2\oplus \rho_{21}^2\,,\\
        &\rho_R = \mathbf{1} \oplus \rho_{11}^3 \oplus \rho_{22}\,.
    \end{aligned}
\end{equation}
The second representation that satisfies our constraints acting crystallographically on bosons acts as $Q_8$ on the left and $S_3\times\Z_2$ on the right. It is generated by
\begin{equation}
    \begin{aligned}
        &r = \left(1,1,i,-i \big{|} -\phi,-\phi^{-1},-\phi,-\phi^{-1}\right) \,,\\
        &s = \left(-1,-1,\SmallMatrix{ 0 & -1 \\
        1 & 0} \big{|}\SmallMatrix{ 0 & 1 \\
        1 & 0},\SmallMatrix{ 0 & 1 \\
        1 & 0}\right)\,.
    \end{aligned}
\end{equation}
and has the same bosonic action of the first one. The fermionic action is in fact obtained from \eqref{eq:dic24sol} by adding an extra $(-1)^{F_R}$ to $r$.

To mass up most twisted sectors, we can consistently add order 2 shifts on the base $S^1$ to both $r$ and $s$. The subgroup of elements without shifts is given by $N=\langle r^2, rs\rangle \simeq Dic_{12}$. In particular, the elements $r^4$ and $r^6$ have trivial actions on left and right movers respectively, so they give could massless gravitini in their twisted sectors. However, they trivially commute with $r$, which projects them out.
$rs$ does not have a trivial action either on the left or on the right.

\subsection{\texorpdfstring{$S_3 \times \Z_4$}{S3 x Z4}}\label{sec:S3Z4}

There is one representation of $S_3\times\Z_4$ implementing our mechanism, acting as $\Z_4\times \Z_2$ on the left and $S_3$ on the right, and generated by
\begin{equation}
    \begin{aligned}
       &r = \left(1,1,1,1 \big{|}\phi,\phi^{-1},\phi,\phi^{-1}\right) \,,\\
        &s = \left(-1,-1,-1,-1 \big{|}\SmallMatrix{     0 & 1 \\
        1 & 0} \SmallMatrix{0 & 1 \\
        1 & 0}\right)\,,\\
        &t = \left(1,1,i,-i \big{|} 1,1,1,1\right)\,,
    \end{aligned}
\end{equation}
where $\phi = e^{2 \pi i/3}$ is a third root of unity.

The bosonic action is a symmetry of lattices \textbf{1-4}, \textbf{12}, \textbf{16} and \textbf{17} in \hyperref[tab:crystals]{Table 2} and it acts as $\Z_4$ on the left and $S_3$ on the right.
For concreteness, we choose a lattice of the form:
\begin{equation}
    \Lambda = \Lambda_1 \oplus \Lambda_R(A_1) \oplus \Lambda_R(A_2) \oplus \Lambda_R(A_2)\,,
\end{equation}
where $\Lambda_1$ is an arbitrary lattice of rank 1. On this lattice, the induced actions on bosons are
\begin{equation}
    \begin{aligned}
        &r = \left(1,1,1,1,1,1 \big{|} 1, 1, 1, 1, \SmallMatrix{0 & -1 \\
        1 & -1}\right)\otimes (-1)^{F_{R}}\,,\\ 
        & s = \left(1,1,1,1,1,1 \big{|} 1, -1, -1, -1, \SmallMatrix{-1 & 1 \\
        0 & 1}\right)\otimes (-1)^{F_{L}}\,,\\
        & t = \left(1,1,\SmallMatrix{0 & 0 & -1 & 0 \\0 & 0 & 0 & -1 \\
    1 & 0 & 0 & 0 \\
    0 & 1 & 0 & 0}\big{|}1,1,1,1,1,1\right)\,.
    \end{aligned}
\end{equation}

We can mass up most twisted sectors by adding order 2 and 4 shift on the base $S^1$ to $s$ and $t$ respectively. The subgroup of element without shifts is given by $N = \langle r,st^2 \rangle \simeq S_3$. The generator $r$ has a trivial action on the left, possibly giving a massless gravitino. However, the central $t$ projects it out since it has a non-trivial left action. The other generator $st^2$ does not act trivially on either the left or the right, so there can be no gravitini in its twisted sector.

\subsection{\texorpdfstring{$G^7_{36}=(\Z_3\times \Z_3) \rtimes \Z_4$}{G736}}\label{sec:G736}

The group of order 36, $G_{36}^7$ has one fermionic solution. It acts as $Dic_{12}$ on the left and $S_3$ on the right. It is generated by
\begin{equation}
    \begin{aligned}
        &a= \left(1,1,\phi,\phi^{-1}\big{|} 1,1,1,1\right)\,,\\
        &b=\left(1,1,1,1\big{|} \phi,\phi^{-1},\phi,\phi^{-1}\right)\,,\\
        &x=\left(-1,-1,\SmallMatrix{0 & 1 \\-1 & 0} \big{|}\SmallMatrix{0 & 1 \\1 & 0},\SmallMatrix{0 & 1 \\1 & 0}\right)\,,
    \end{aligned}
\end{equation}
where $\phi = e^{2 \pi i/3}$ is a third root of unity.

The bosonic action is a symmetry of lattice \textbf{12} in \hyperref[tab:crystals]{Table 2}, acting as $Dic_{12}$ on the left and as $S_3$ on the right.
On this lattice, we can write the left moving action as

\begin{equation}
    \begin{aligned}
        a &=\left(\SmallMatrix{0 & -1 \\
        1 & -1},\SmallMatrix{0 & -1 \\
        1 & -1},1,1\big{|}1,1,1,1,1,1\right)\,, \\
        b &=\left(1,1,1,1,1,1 \big{|} \SmallMatrix{0 & -1 \\
        1 & -1},1,1,1,1\right) \otimes (-1)^{F_R}\\ 
        x &=\left(\SmallMatrix{ 0 & 0 & 1 & 0 \\ 0 & 0 & 0 & 1 \\ -1 & 0 & 0 & 0 \\ 0 & -1 & 0 & 0},1,1 \big{|} \SmallMatrix{-1 & 1 \\
        0 & 1},-1,-1,-1,1\right) \otimes (-1)^{F_L}
    \end{aligned}
\end{equation}

To mass up some twisted sectors we can put a shift of order 4 on the $S^1$ base circle in $x$. The elements without shifts are those in the commutator subgroup $N =[G_{36}^7,G_{36}^7] = \langle a, b\rangle \simeq \Z_3\times\Z_3 $.  The two $\Z_3$ generators, $a$ and $b$, have a trivial action on the right and left movers, respectively, but they act non-trivially on the other side. Since they commute, they project out each other's potential twisted sector gravitino.

\subsection{\texorpdfstring{$S_3 \times \Z_6$}{S3 x Z6}}\label{sec:S3Z6}

The group $S_3 \times \Z_6$ has one fermionic representation that implements one-loop vacuum energy cancellation without supersymmetry. It acts as $\Z_6 \times \Z_2$ on left movers and as $S_3$ on right movers. It is generated by
\begin{equation}
    \begin{aligned}
       &r = \left(1,1,1,1 \big{|}\phi,\phi^{-1},\phi,\phi^{-1}\right) \,,\\
        &s = \left(-1,-1, 1, 1\big{|}\SmallMatrix{0 & 1 \\1 & 0} \SmallMatrix{0 & 1 \\1 & 0}\right)\,,\\
        &t = \left(1,1,\zeta^{-1},\zeta \big{|} 1,1,1,1\right)\,,
    \end{aligned}
\end{equation}
with $\phi = e^{2 \pi i/3}$ is a third root of unity and $\zeta = e^{2\pi i/6}$ a sixth root of unity.

The bosonic action is a symmetry of lattice \textbf{12} in \hyperref[tab:crystals]{Table 2} and acts as $\Z_6$ on the left and $S_3$ on the right. 
On this lattice, the bosonic action can be written as:
\begin{equation}
    \begin{aligned}
        &r = \left(1,1,1,1,1,1\big{|}1,1,1,1,\SmallMatrix{0 & -1 \\1 & -1}\right)\otimes (-1)^{F_{R}}\,,\\
        &s = \left(1,1,- 1,- 1,- 1,- 1\big{|}1,-1,-1,-1,\SmallMatrix{-1 & 1 \\
        0 & 1}\right)\,,\\
        & t =\left(1,1,\SmallMatrix{1 & -1 \\1 & 0},\SmallMatrix{1 & -1 \\1 & 0}\big{|}1,1,1,1,1,1\right)\,.
    \end{aligned}
\end{equation}
We mass up most twisted sector by adding order 2 and 6 shifts on the base circle to $s$ and $t$ respectively. The subgroup of elements without shifts is $N =  \langle r,st^3 \rangle\simeq S_3$. The generator $r$ acts trivially on the left. However, potential gravitini are projected out by the central $t$, whose left action is non-trivial. The generator $st^3$ acts non-trivially both on the left and on the right, so it cannot give massless gravitini.

\subsection{\texorpdfstring{$S_3 \times Q_8$}{S3 x Q8}}\label{sec:S3Q8}
The group $S_3\times Q_8$ has one fermionic representation, acting as $\Z_2 \times Q_8$ on the left and $S_3$ on the right. The generators are:
\begin{equation}
    \begin{aligned}
    &i=\left(1,1,\SmallMatrix{0 & -1 \\1 & 0}\big{|}1,1,1,1\right) \,,\\
    &j = \left(1,1,i,-i\big{|}1,1,1,1\right) \,,\\
    &r = \left(1,1,1,1 \big{|}\phi,\phi^{-1},\phi,\phi^{-1}\right) \,,\\
    &s = \left(-1,-1, 1, 1\big{|}\SmallMatrix{0 & 1 \\1 & 0} \SmallMatrix{0 & 1 \\1 & 0}\right)\,,\\
    \end{aligned}
\end{equation}
where $\phi = e^{2 \pi i/3}$ is a third root of unity.
The bosonic representation acts as a $Q_8$ on left movers and $S_3$ on right movers and is a symmetry of lattices \textbf{1-4} in \hyperref[tab:crystals]{Table 2}. The action on the left decomposes as\footnote{\label{fn:tensor_product_rep} Since representations of a direct product group $G = G_1 \times G_2$ are given by tensor products of irreducible representations of the individual factors we denote $\rho_1\otimes \rho_2=: (\rho_1,\rho_2)$}
\begin{equation}
        \rho_L = (\mathbf{1},\mathbf{1})^{\oplus 2} \oplus (\mathbf{1},\tau)^{\oplus 2}\,,
\end{equation}
where $\tau$ is the faithful irreducible representation of $Q_8$ defined in \hyperref[sec:Q8_anomaly]{Section 4.1}, while the representation on the right is given by
\begin{equation}
     \rho_R = (\mathbf{1},\mathbf{1}) \oplus (\mathbf{sign},\mathbf{1})^{\oplus 3} \oplus (\mathbf{std},\mathbf{1}) \,.
\end{equation}
Here, $\mathbf{sign}$ and $\mathbf{std}$ are again the sign and standard irreducible representations of $S_3$, respectively.

We mass up twisted sectors by adding order 2 shifts on the base $S^1$ to $i,j$ and $s$.
The subgroup with trivial shifts is given by $N = \langle r,si, sj \rangle \simeq Dic_{24}$.
In $N$, the only elements acting trivially on the left are powers of $r$. $r$ commutes with the central $i^2$, so any potential gravitino is projected out.

The only element in $N$ acting trivially on the right is $i^2$, which is central, so its associated potential gravitino is projected out.

\subsection{\texorpdfstring{$S_3 \times Dic_{12}$}{S3 x Dic12}}\label{sec:S3Dic12}

The group $S_3\times Dic_{12}$ has one fermionic solution. The fermionic action is $Dic_{12}\times \Z_2$ on the left and $S_3$ on the right, generated by:
\begin{equation}
    \begin{aligned}
    &a=\left(1,1,\zeta^{-1},\zeta\big{|}1,1,1,1\right) \,,\\
    &x= \left(1,1,\SmallMatrix{0 & -1 \\1 & 0} \big{|}1,1,1,1\right) \,,\\
    &r = \left(1,1,1,1 \big{|}\phi,\phi^{-1},\phi,\phi^{-1}\right) \,,\\
    &s = \left(-1,-1, 1, 1\big{|}\SmallMatrix{0 & 1 \\1 & 0} \SmallMatrix{0 & 1 \\1 & 0}\right)\,,\\
    \end{aligned}
\end{equation}
where $\phi = e^{2 \pi i/3}$ is a third root of unity and $\zeta = e^{2\pi i/6}$ a sixth root of unity.
The bosonic action is a symmetry of lattice \textbf{12} in \hyperref[tab:crystals]{Table 2}, it acts as $Dic_{12}$ on the left movers and $S_3$ on the right movers. On this lattice the bosonic action is:
\begin{equation}
    \begin{aligned}
    &a=\left(1,1,\SmallMatrix{1 & -1 \\ 1 & 0},\SmallMatrix{0 & 1 \\ -1 & 1}\big{|}1,1,1,1,1,1\right) \,,\\
    &x = \left(1,1,\SmallMatrix{0 & 0 & 1 & 0 \\0 & 0 & 0 & 1 \\
    -1 & 0 & 0 & 0 \\
    0 & -1 & 0 & 0}\big{|}1,1,1,1,1,1\right) \,,\\
    &r = \left(1,1,1,1,1,1\big{|}1,1,1,1, \SmallMatrix{ 0 & -1 \\ 1 & -1}\right) \otimes (-1)^{F_R} \,,\\
    &s = \left(1,1,- 1,- 1,- 1,- 1\big{|}1,-1,-1,-1,\SmallMatrix{-1 & 1 \\ 0 & 1}\right)\,.\\
    \end{aligned}
\end{equation}

To lift masses of twisted sectors, we put an order 4 shift to $x$ and order 2 shifts on $a$ and $s$, all acting on the base $S^1$. The elements without shift are those in $N= \langle r, a^2, as \rangle \simeq S_3 \times \Z_3$.
The generator $r$ acts trivially on the left, but commutes with the $Dic_{12}$ factor, which projects out any potential gravitino. $a^2$ acts trivially on the right, but it commutes with the $S_3$ factor, which projects out any potential gravitino. Finally, $as$ does not act trivially on either side.

\subsection{\texorpdfstring{$S_3 \times SL(2,3)$}{S3 x SL(2,3)}}\label{sec:S3SL23}
Finally, also the order 144 group $S_3\times SL(2,3)$, implements the non-supersymmetric cancellation of the one-loop vacuum energy. The fermionic representation acts as $SL(2,3)\times\Z_2$ on the left and $S_3$ on the right. It is generated by
\begin{equation}
    \begin{aligned}
    &i=\left(1,1,\SmallMatrix{i & 0 \\ 0 & -i}\big{|}1,1,1,1\right) \,,\\
    &j = \left(1,1,\SmallMatrix{0 & 1 \\-1 & 0}\big{|}1,1,1,1\right) \,,\\
    &t = \left(1,1,\tfrac{1}{2}\SmallMatrix{-1+i & 1+i \\ -1+i & -1-i}\big{|}1,1,1,1\right) \,,\\
    &r = \left(1,1,1,1\big{|}\SmallMatrix{-1 & -1 \\1 & 0} \SmallMatrix{-1 & -1 \\1 & 0}\right)\,,\\
    &s = \left(-1,-1,1,1\big{|}\SmallMatrix{-1 & -1 \\0 & 1} \SmallMatrix{-1 & -1 \\0 & 1}\right)\,,\\
    \end{aligned}
\end{equation}
The induced bosonic action is an $SL(2,3)$ on left movers and $S_3$ on right movers and represents a symmetry of lattice \textbf{4} in \hyperref[tab:crystals]{Table 2}. In particular, we find
\begin{equation}
    \rho_L = (\mathbf{1},\mathbf{1})^{\oplus 2} \oplus (\mathbf{1},\rho_2)^{\oplus 2}
\end{equation}
for left movers and
\begin{equation}
    \rho_R = (\mathbf{1},\mathbf{1}) \oplus (\mathbf{sign},\mathbf{1})^{\oplus 3} \oplus (\mathbf{std},\mathbf{1})
\end{equation}
for right movers.

As usual, to mass up most twisted sectors, we add order 2 and 3 shifts on the base $S^1$ to $s$ and $t$ respectively. Elements without shifts are all in the commutator subgroup, given by $[S_3 \times SL(2,3),S_3 \times SL(2,3)]\simeq \Z_3\times Q_8 = \langle r, i, j  \rangle$. The generators $i$ and $j$ commute with $r$ and the first two have a non-trivial action where the other acts trivially, and vice versa, so gravitini are projected out.

\subsection{Decompactification limits}

Since we have built orbifolds whose action on the base $S^1$ is purely geometric, its radius $r$ is a modulus. It is then natural to investigate what happens in the $r\rightarrow\infty$ decompactification limit. To do so, we need to look at the action of space group elements which contain shifts in this regime.

Consider an element $g=(\theta,v)$ of the space group, with the shift $v = r/k$ of order $k$ only along the base $S^1$. Since its action on the circle is purely geometric we can equivalently view this quotient as a holonomy of $\theta$ acting on the internal fields when going around a smaller circle of radius $r/k$, see e.g. \cite{Dabholkar:2005ve}. We are in this sense ``compactifying" on a non-trivial fibration over the base $S^1$, with holonomy specified by $\theta$. Then in the $r\rightarrow \infty$ decompactification limit the fibration trivializes, since the base space is now $\R$. From the orbifold point of view the action of the element $g$ becomes trivial in the limit.
After decompactifying, only those elements of the space group $G$ which act trivially on the circle keep a non-trivial action.

\begin{figure}[t]
    \centering
    \includegraphics[width=0.8\textwidth]{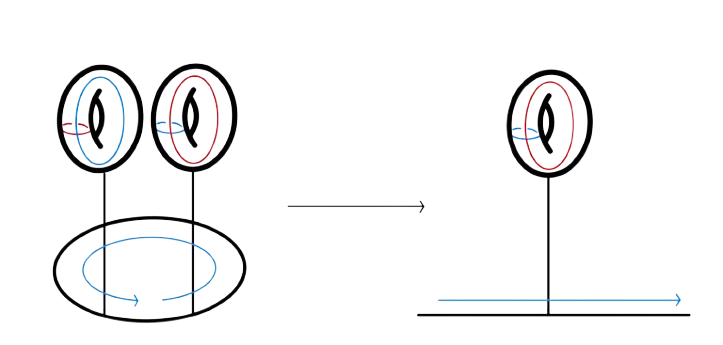}
    \caption{A pictorial representation of the decompactification limit for an element $g = (\theta,v)$ with a shift $v$ along the base. Here, as an example, $\theta$ acts by exchanging the two fundamental cycles of the torus. On the left, we view this as a holonomy of $\theta$ around the circle $S^i/v$. The decompactification limit on the right can be interpreted as a trivialization of the associated fibration, where $\theta$ does not act anymore.}
    \label{fig:decompactify}
\end{figure}

In our orbifolds these elements are precisely those of the subgroup of elements without shifts, $N$, which we already computed above to check for twisted sector gravitini. In the limit we are therefore considering an asymmetric $T^5/N$ compactification, and we can study it with the same tools.

In particular, we can quickly read off the amount of supersymmetry restored in the limit, at least in the untwisted sector, since we already have the action of $N$ on supercharges. We report the results in \hyperref[tab:decompactified_models]{Table 4}.

In all cases where $N$ is cyclic, it automatically leads to a supersymmetric theory, since by condition \textbf{III.} every element in $G$ preserves at least one supercharge. Then if $N\simeq\Z_k$ the whole group preserves the same supercharge, and the decompactified theory is supersymmetric.

Less trivially so, this is the case also for another set of examples: $D_{12}$,  $G_{36}^7$ and $SL(2,3) \times S_3$. All generators of the respective subgroup $N$ in fact preserve some common supercharges, so again the untwisted sector already gives spacetime supersymmetry.

Much more interesting are the $G= Dic_{24},S_3 \times \Z_6, S_3 \times Q_8$, and $S_3\times Dic_{12}$ cases. In these models SUSY is explicitly broken in the untwisted sector. Through the same methods as the main text, we can check for potential massless gravitini in the twisted sectors. The only difference is that when looking at the $g$-twisted sector, only projections by elements of the subgroup $N$ are allowed.

For the $G=S_3 \times \Z_6$ model, we find $N = S_3$, and the generator $r$ acts trivially on the left. Since its centralizer in $N$ is $\mathcal{C}_N(r) = \langle r \rangle$, the massless gravitino in the $r$-twisted sector survives the orbifold projection, and SUSY is restored in the twisted sector.

In the $G=Dic_{24}, S_3 \times Q_8, S_3 \times Dic_{12}$ models, there are no twisted gravitini and therefore SUSY is broken even in the decompactification limit. Note that the vacuum energy is still vanishing in the limit. The only thing that could spoil its vanishing would be if some pair of elements who did not commute in $G$, but commute in $N$, would show up in the orbifold sum \eqref{eq:orbifold_partition_function}. This can never happen since $N$ contains the commutator subgroup $[G,G]$, so no non-trivial commutation relations are trivialized in the limit.

In fact, one can see at a glance from \hyperref[tab:decompactified_models]{Table 4} that the groups $N$ for which there is no SUSY in the limit appear already as full orbifold group $G$ in our list of solutions. One can explicitly check that the representations are also exactly the same, as expected, so these solutions decompactify to other solutions in our list.

Since in the limit the theory is genuinely 5-dimensional, this means that the shifts on the $S^1$ were not really needed in these models, and the point group was enough to guarantee a vanishing vacuum energy. 

In the $G=D_{12}$, $N= S_3$ case, SUSY restoration in the twisted sector can be avoided by adding a shift exactly as in the $G=S_3$ example of \hyperref[sec:S3]{Section 5.1}.

In the $Q_8$ example of \hyperref[sec:Q8]{Section 5.2}, we have a purely geometric shift action on a full $T^2$ instead of only an $S^1$. We could therefore let $i$ and $j$ act with shifts on both circles, then decompactify both dimensions in the same way leading to a $T^4/\Z_4$ supersymmetric orbifold. Alternatively, we can add shifts to the generators $i$ and $j$ on two different $S^1$'s. We could then decompactify along only one direction to get a $T^5/\Z_4$ supersymmetric orbifold in which only either $i$ or $j$ acts trivially, or decompactify both directions to get a $T^4/\Z_2$ supersymmetric orbifold.

Some orbifolds are identified in the decompactification limits. As can be checked directly with the generators given earlier, the $G=S_3,S_3 \times \Z_3$ and the first of the $G=Dic_{12}$ examples decompactify to the same supersymmetric $\Z_3$ asymmetric orbifold. Similarly, the $G= D_{12}$ and $G=S_3 \times \Z_4$ models decompactify to the same $N=S_3$ supersymmetric orbifold. 

\begin{table}[!ht]\label{tab:decompactified_models}
\renewcommand{\arraystretch}{1.3}\centering
\begin{tabular}{|
>{$}c<{$}|
c|
c|
>{$}c<{$}|
>{$}c<{$}|
} \hline
 \text{Orbifold group $G$} & $G_{\text{dec}} = N$ & $n^{\text{untw}}_Q$ & N_F & N_B  \\ \hline
 \hyperref[sec:S3]{S_3} & $\Z_3$ & 16 &  1 \oplus \Z_3 & 1 \oplus \Z_3  \\ \hline
 \text{\multirow{3}{*}{\hyperref[sec:Q8]{$Q_8$}}} & \multirow{2}{*}{$\Z_4$} & 8 &  \Z_2 \oplus \Z_4 &  \Z_2 \oplus \Z_4  \\
 & & 8 &  \Z_2 \oplus \Z_4  &  1 \oplus \Z_4  \\ \cline{2-5}
& $\Z_2$ & 24 &  1 \oplus \Z_2  &  1 \oplus \Z_2  \\
 \hline
 \text{\multirow{2}{*}{\hyperref[sec:Dic12]{$Dic_{12}$}}} & \multirow{2}{*}{$\Z_3$} & 16 &  1 \oplus \Z_3 &  1 \oplus \Z_3  \\ 
 && 8 &  \Z_3\oplus \Z_3 &  \Z_3\oplus \Z_3 \\ \hline
 \hyperref[sec:D12]{Dic_{12}} & $S_3$ & 8 &  \Z_2 \oplus S_3 &  \Z_2 \oplus S_3 \\ \hline
 \hyperref[sec:S3Z3]{S_3 \times  \Z_3} & $\Z_3$ & 16 &  1 \oplus \Z_3 &  1 \oplus \Z_3 \\ \hline
 \hyperref[sec:Dic24]{Dic_{24}} & \hyperref[sec:Dic12]{$Dic_{12}$} & 0 &  \Z_4 \oplus S_3&  \Z_4 \oplus S_3 \\ \hline
 \hyperref[sec:S3Z4]{S_3 \times  \Z_4} & $S_3$& 8 &  \Z_2 \oplus S_3 &  \Z_2 \oplus S_3 \\ \hline
 \hyperref[sec:G736]{G^7_{36}} & $\Z_3 \times \Z_3$& 8 &  \Z_3 \oplus \Z_3&  \Z_3 \oplus \Z_3 \\\hline
 \hyperref[sec:S3Z6]{S_3 \times  \Z_6} & \hyperref[sec:S3]{$S_3$} & 0 &  \Z_2 \oplus S_3 &  1 \oplus S_3 \\ \hline
 \hyperref[sec:S3Q8]{S_3 \times  Q_8} & \hyperref[sec:Dic24]{$Dic_{24}$} & 0  &  Q_8 \oplus S_3   &  Q_8 \oplus S_3 \\ \hline
 \hyperref[sec:S3Dic12]{S_3 \times  Dic_{12}} & \hyperref[sec:S3Z3]{$S_3 \times \Z_3$}& 0  &  \Z_6 \oplus S_3 &  \Z_3 \oplus S_3 \\ \hline
 \hyperref[sec:S3SL23]{S_3 \times SL(2,3)} & $\Z_3\times Q_8$ & 8  &  Q_8 \oplus \Z_3&  Q_8 \oplus \Z_3 \\ \hline
\end{tabular}
\caption{Decompactification limits of our $T^5 \times S^1$ orbifolds. Here, $G_{\text{dec}}$ is the orbifold group in the decompactification limit, given by the subgroup $N$ of elements of $G$ without shifts. It acts on the full $T^5$ in all cases except for $Q_8$ which acts on a $T^4$. In this case, different choices of shifts lead to different $N$, as discussed above. $n^{\text{untw}}_Q$ is the number of conserved supercharges in the untwisted sector.}
\end{table}

Finally, one can also consider the opposite limit, where $r\rightarrow 0$. The natural way to do that is by $T$-dualizing along the base circle and taking the decompactification limit in the dual coordinate $\tilde{r} \sim 1/r$. An order $k$ shift in the $r$ coordinate $T$-dualizes to an order $k$ shift in the dual coordinate $\tilde{r}$ \cite{Dabholkar:2005ve}, while the twist in the ``fiber" remains untouched since the $T^5$ is orthogonal to the base $S^1$. We conclude that the $r\rightarrow 0$ limit gives the same theory and the considerations about SUSY restoration equally apply.

\section{Conclusions and outlook}

In this work, we took a step towards the classification of non-supersymmetric orbifolds with non-abelian point group and vanishing one-loop vacuum energy.

This extends the results of \cite{Larotonda:2026hxy}, which focused on abelian point groups, to non-abelian orbifold point groups within a restricted class. The more rigid nature of the representation theory of non-abelian groups leads to a much smaller number of solutions, down from the 2000 or so abelian solutions of \cite{Larotonda:2026hxy} to only 17 of them, at least in our restricted class.

Since we always have as a modulus the radius of the geometric base $S^1$, we have explored their decompactification limits, showing that all our solutions can alternatively be viewed as special Scherk-Schwarz-like compactifications of some $T^5$ asymmetric orbifold. Most of these limits restore SUSY, but some of them are genuinely non-supersymmetric vacua with no tachyons.

Along the way we have developed tools to study in detail some aspects of the relatively unexplored landscape of non-abelian toroidal orbifolds, focusing on the cancellation of global anomalies, even beyond the classic one-loop conditions, and the counting of supercharges, which is apparent when working directly with fermionic representations.
We expect these techniques to be useful beyond the specific scope of the models studied here.

The natural direction in which to extend this work is of course to drop the restriction on the $T^5 \times S^1$ and explore point groups acting faithfully on the whole $T^6$. While on one hand there are many more candidate groups to look at in the first place, we expect the absence of the geometric shift on the base $S^1$ to lead to accidental SUSY restoration in many cases. We hope to come back to the full classification soon.

The biggest open question remains the same as in \cite{Larotonda:2026hxy}: for which of these models, if any at all, does the cancellation persist at higher loops? In this sense the non-abelian orbifolds of this work might be optimal candidates.

While, at least in the toroidal context, it is hard to imagine the cancellation surviving at sufficiently high order to provide a realistic suppression of $\Lambda \sim 10^{-120}$, finding models with higher loop vanishing has interesting theoretical consequences. The breaking of a conservation law at higher loops in string perturbation theory is reminiscent of what happens with global internal symmetries on the worldsheet \cite{Heckman:2024obe}. If one takes the point of view that supersymmetry is what protects the vanishing of the vacuum energy, this seems to point towards some non-invertible version of SUSY. This concept has, to the authors' knowledge, not been formalized in the literature. We believe, however, that should such a thing as non-invertible supersymmetry exist, the orbifolds we present are prime candidates in which to look for it.

Understanding this mechanism in full generality may allow for it to be applied to more general orbifold CFTs, and eventually lead to solutions with realistic values for the cosmological constant.

\section*{Acknowledgments}

We would like to thank Bruno Bento, Matilda Delgado, Iñaki García-Etxebarría, Héctor Parra de Freitas, Yuji Satoh and Ethan Torres for enlightening discussions. We are especially grateful to Miguel Montero for valuable comments on the draft. V.L. would like to thank Kavli IPMU for their hospitality during the early stages of this work.
B.F. is supported by a Juan de la Cierva contract (JDC2023-050850-I) from Spain’s Ministry of Science, Innovation and Universities.
M.T. is supported by the FPI grant PRE2022-102286 from Spanish National Research Agency from the Ministry of Science and Innovation. The work of B.F. and M.T. is also supported through the grants CEX2020-001007-S, PID2021-123017NB-I00 and PID2024-156043NB-I00, funded by MCIN/AEI/10.13039/501100011033, and ERDF, EU.

\appendix

\section{ Review of basics of lattices and crystals}\label{app:cyclotomic}

We review here some basic facts about the theory of lattices and crystals, in particular in relation to their implementation in GAP.

A \textit{lattice} is a free $\Z$-module $\Lambda$, together with a choice of quadratic form $q$. One can always think of this as an embedding $\Z^d\hookrightarrow \R^d$, with the quadratic form $q$ given by restriction of the canonical metric on $\R^d$. A \textit{lattice automorphism} is a $\Z$-module automorphism of $\Lambda$ which preserves the quadratic form. From the point of view of the embedding, this is a rotation $\theta \in O(d)$ which preserves the lattice. In a proper basis, it can always be regarded as an element $\theta \in O(d)\cap GL(d;\Z)$. We refer to the group of lattice automorphisms in this sense as the \textit{point group}.

Most of the mathematics literature on the classification of lattices uses notation historically derived from chemistry. This is precisely the terminology CARAT \cite{Plesken:zm0025} uses, so we review it here. Much more detail can be found e.g.~in \cite{Opgenorth:js0065}.
Given a quadratic form $q$ and its associated matrix $Q$, its \textit{Bravais group} $B(q)$ is defined as the stabilizer of $Q$ in $GL(d;\Z)$.

A \textit{Bravais lattice} is an equivalence class of lattices, where we identify two lattices if they have the same automorphism group in the sense above, up to conjugation in $GL(d,\Z)$. Since under this equivalence the entire information is captured by the automorphism group, one can describe each equivalent class as the Bravais group of the quadratic form associated to the lattice. Equivalently, it can be described as a subgroup of $GL(d;\Z)$, up to conjugation in $GL(d;\Z)$. Most of the classification results are in fact dedicated to studying equivalence classes of subgroups of $GL(d;\Z)$ rather than lattices themselves as defined abstractly above.

A useful equivalence relation between subgroups of $GL(d;\Z)$ is the following. Take two subgroups $G$ and $H$, and consider the space of quadratic forms fixed by either, call them $\mathcal{F}(G)$ and $\mathcal{F}(H)$ respectively. $G$ and $H$ are said to be \textit{Bravais equivalent} if the Bravais groups $B(G) := B(\mathcal{F}(G))$ and $B(H):=B(\mathcal{F}(H))$ are equivalent up to $GL(d;\Z)$ conjugation.
Note that since in general $B(G)$ is bigger than $G$, this can relate two groups $G$,$H$ of different orders. Furthermore, in a Bravais lattice class, all lattices are Bravais equivalent.

Finally, a closely related notion is that of a \textit{crystal family}, which is defined as a coarser equivalence class of subgroups of $GL(d;\Z)$.

Two groups $G$ and $H$ are said to be in the same crystal family if there is a finite sequence of subgroups $\{G_i\}$ of $GL(d;\Z)$ such that $G_0=G$, $G_n = H$, and at each step $G_i$ and $G_{i+1}$ are either conjugate over $GL(d;\Q)$ or Bravais equivalent in the sense just defined.

As an example, the \textit{cubic} crystal family in $d=3$ contains three different inequivalent lattices: the cubic $\Z^3$, but also the face centered cubic (FCC) lattice and the body centered cubic lattice (BCC). These three are Bravais equivalent to, respectively, $\Lambda_R(B_3),\Lambda_R(D_3),\Lambda_W(D_3)$, and we will refer to them as such in the main text.

CARAT generally uses crystal families as its main classification tool, but the equivalence relation we care about is that of Bravais groups. This is essentially because in asymmetric orbifolds all the radii are automatically fixed at the self-dual value, so the Bravais group fully specifies the enhanced symmetry point in Narain moduli space.
Therefore, in our algorithm, the condition we impose for the symmetry to be crystallographic is that both the left- and right-moving bosonic actions land in the same five-dimensional Bravais group. 

\subsection{The gluing construction}

We describe here a way to construct Narain lattices with enhanced symmetries, following the notation in \cite{Baykara:2024vss}. Let $\Lambda$ be an even integral lattice. Then necessarily it is contained in its dual, $\Lambda \subset \Lambda^*$. The failure of this inclusion to be an isomorphism is measured by the \textit{discriminant group}
\begin{equation}
    \mathcal{D}(\Lambda) = \Lambda^*/\Lambda\,,
\end{equation}
which is a finite abelian group. It inherits the quadratic form $q$ from $\Lambda$, but only mod 2:
\begin{equation}
    \begin{aligned}
        \bar{q}:\mathcal{D}(\Lambda) &\rightarrow \Q/2\Z\,,\\
                \bar{q}([x]) &= q(x) \quad \text{mod 2}\,.
    \end{aligned}
\end{equation}
The discriminant groups of two lattices $\Lambda_1$ and $\Lambda_2$ are called isometric if there exists a map 
\begin{equation}
    \begin{aligned}
        f:\mathcal{D}(\Lambda_1) \rightarrow \mathcal{D}(\Lambda_2)\,,
    \end{aligned}
\end{equation}
such that the following commutes:
\begin{equation}
    \begin{tikzcd}
        \mathcal{D}(\Lambda_1) \arrow[rr, "f"] \arrow[dr, "\bar{q}_1"'] & & \mathcal{D}(\Lambda_2) \arrow[dl, "\bar{q}_2"] \\
        & \mathbb{Q}/2\mathbb{Z} &
    \end{tikzcd}\,.
\end{equation}
Now let $\Lambda_1$ and $\Lambda_2$ be two even lattices of rank $d$, and $f$ be an isometry between their discriminant groups. Then we can build an even self-dual lattice of signature $(d,d)$ as 
\begin{equation}
    \Gamma^{d,d}(\Lambda_1,\Lambda_2) = \left\{ (x,y) \in \Lambda_1^* \oplus \Lambda_2^* | f([x]) = [y] \right\}\,,
\end{equation}
with the quadratic form given by
\begin{equation}\label{eq:glueQ}
    q(x,y)= q_1(x)-q_2(y)\,.
\end{equation}

In our cases, we will take $\Lambda_1=\Lambda_2$, with the isometry given trivially by the identity. This recovers the construction given in \eqref{eq:extraSymm}. Replacing \eqref{eq:glueQ} with $q_1(x) + q_2(x)$ gives a not necessarily even, self-dual, lattice of signature $(2d,0)$.

\subsection{Eigenvalues and cyclotomic polynomials}

Symmetries of lattices are very constrained in terms of their eigenvalues. These can be studied by using cyclotomic polynomials. See \cite{Baykara:2024vss,Baykara:2025gcc} for recent discussions in the physics literature. We summarize the result here briefly, for more details see \cite{SzczepanskiAndrzej2012Gocg,Kuzmanovich01022002}. Consider a rank $n$ lattice $\Lambda$ and an automorphism $g \in \text{Aut}(\Lambda)$. Choosing a lattice basis, this is represented by a $n \times n$ integral matrix $G \in GL(n;\Z)$. Its eigenvalues are found as roots of the minimal polynomial of this matrix, call it $\mu_G$. One of its basic properties is that it divides all polynomials $P$ that annihilate $G$, i.e.~those for which $P(G)=0$.
\\For an order $k$ automorphism $g$, its matrix is  annihilated by $P(x)=x^k-1$.

The $k$-th cyclotomic polynomial $\Phi_k(x)$ is precisely defined as the unique irreducible polynomial with integer coefficients which divides $x^k-1$ but does not divide $x^{k'}-1$ for any $k'<k$. This has the consequence that 
\begin{equation}
    x^k - 1 = \prod_{d|k} \Phi_d(x)\,.
\end{equation}
For prime $k$ this is particularly easy, and reads
\begin{equation}
    x^k-1 = (x-1) \Phi_k(x)\,.
\end{equation}
Since both factors are irreducible, one of them has to be $\mu_G$. This means that the eigenvalues of an order $k$ automorphism are either all $1$, or they are found as the roots of the cyclotomic polynomial $\Phi_k$.

The first few cyclotomic polynomials read as follows:
\begin{equation}   \label{eq:cyclotomic}
    \begin{aligned}
        \Phi_1(x) &= x - 1 \\
        \Phi_2(x) &= x + 1 \\
        \Phi_3(x) &= x^2 + x + 1 \\
        \Phi_4(x) &= x^2 + 1 \\
        \Phi_5(x) &= x^4 + x^3 + x^2 + x + 1 \\
        \Phi_6(x) &= x^2 - x + 1 \\
        \Phi_7(x) &= x^6 + x^5 + x^4 + x^3 + x^2 + x + 1 \\
        \Phi_8(x) &= x^4 + 1 \\
        \Phi_9(x) &= x^6 + x^3 + 1 \\
        \Phi_{10}(x) &= x^4 - x^3 + x^2 - x + 1\,.
    \end{aligned}
\end{equation}
From these one can read off our cases of interest. In particular, for $\Z_5$ actions we see that the symmetry must act either trivially or on a four-dimensional block, with eigenvalues
\begin{equation}
    (\omega,\omega^2,\omega^3,\omega^4) \qquad \text{for} \qquad \omega = e^{2 \pi i/5} \,.
\end{equation}

\section{Group extensions}\label{app:ext}

We review here some basic facts about group extensions. For a detailed discussion see \cite{brown1982cohomology}.

Given a group $G$ and an abelian group $A$, an extension $E$ of $G$ by $A$ is an exact sequence of groups
\begin{equation}
    0 \rightarrow A \xrightarrow[]{i} E \xrightarrow[]{\pi}G \rightarrow 0\,.
\end{equation}
We will be interested in central extensions, in which the image of $A$ lies in the center of $E$. From now on all extensions we write are central.

If there is a homomorphism $s:G\rightarrow E$ such that $\pi s= id_G$ then the sequence splits, and it is equivalent to
\begin{equation}
    0 \rightarrow A \xrightarrow[]{i} A \times G \xrightarrow[]{\pi}G \rightarrow 0\,.
\end{equation}
The set of inequivalent non-trivial extensions of $G$ by $A$, call it $\mathcal{E}(G,A)$, is in one to one correspondence with the group cohomology $H^2(G;A)$. This is the main statement of Theorem 3.2 of \cite{brown1982cohomology}. A corollary of the same theorem is that group extensions are functorial in the following sense:
\begin{itemize}
    \item[1.] $\mathcal{E}(G,\,\cdot \,):\underline{\text{Ab}}\rightarrow \underline{\text{Set}}$ is a covariant functor.
    \item[2.] $\mathcal{E}(\,\cdot \,,A):\underline{\text{Grp}}\rightarrow \underline{\text{Set}}$ is a contravariant functor.
\end{itemize}
We are interested in using the second property, as follows.

We start from a bosonic representation $\rho_B:G_B\rightarrow SO(6)$, and we want to lift it through the canonical double cover $\pi:Spin(6)\rightarrow SO(6)$, which defines a $\Z_2$ central extension. Because of functoriality, we can write the following commutative diagram defining a fermionic representation $\rho_F$:

\begin{equation}\label{eq:pullback}
    \begin{tikzcd}
0 \arrow[r] & \mathbb{Z}_2 \arrow[r] \arrow[d, equal] & G_F \arrow[r] \arrow[d,"\rho_F"] & G_B \arrow[r] \arrow[d,"\rho_B"] & 0 \\
0 \arrow[r] & \mathbb{Z}_2 \arrow[r] & \text{Spin}(N) \arrow[r, "\pi"] & \text{SO}(N) \arrow[r] & 0
\end{tikzcd}\,.
\end{equation}

Now there are two possibilities: the sequence either splits or it does not, depending on $\rho_B$.
The obstruction is given by the second Stiefel-Whitney class of the representation
\begin{equation}
    w_2(\rho_B) \in H^2(G_B;\Z_2),
\end{equation}
as reviewed below.
If this vanishes, the extension is trivial:
\begin{equation}
    G_F \simeq G_B \times \Z_2\,.
\end{equation}
Since the vanishing of $w_2(\rho_B)$ is required for the cancellation of global anomalies \cite{Freed:1987qk}, this justifies the restriction to trivial extensions in the main text.

\subsection{Stiefel-Whitney classes and group extensions}\label{app:char}

Usually characteristic classes are algebraic invariants associated to fiber bundles. For a finite group $G$, there is a well-defined notion of a characteristic class of a representation $G \rightarrow SO(d)$, which is built as follows.

From a representation $\rho: G \rightarrow SO(d)$ one can naturally build a vector bundle, called the associated bundle, over the classifying space $BG$ as the quotient:
\begin{equation}
    E_{\rho} = EG \times_\rho \R^{d} := (EG \times \R^{d})/ \sim \,,
\end{equation}
where the equivalence is built through the given representation $\rho$ acting on $R^d$ and the natural free action of $G$ on its universal bundle $EG$:
\begin{equation}
    (e\cdot g ,v ) \sim (e,\rho(g)\cdot v)\,.
\end{equation}
The projection inherited from $EG$ under this quotient turns $E_\rho$ into a flat bundle on $BG$, with fiber $\R^d$ and holonomy given by $\rho(G) \subset SO(d)$. This is because the classifying space $BG$ is an Eilenberg-MacLane space $K(G;1)$, so it has $\pi_1(BG)\simeq G$ and all higher homotopy groups vanishing.

In fact the mapping between a representation into $SO(d)$ and its associated bundle has an inverse: given a flat bundle over $BG$, its holonomy is by definition a map $G\rightarrow SO(d)$, which defines a $\rho$.

Characteristic classes of the representation $\rho$ are defined as the characteristic classes of its associated bundle over $BG$. For us the relevant one will be the second Stiefel-Whitney class
\begin{equation}
    w_2(\rho) := w_2(E_\rho) \in H^2(BG;\Z_2) \simeq H^2(G;\Z_2)\,.
\end{equation}
In this context the interpretation of $w_2$ is the usual one: it is the obstruction to the existence of a spin structure on $E_\rho$, or the possibility of lifting $E_\rho$ to a $Spin(d)$ bundle. More precisely, if $w_2(\rho)=0$ then there exists a $Spin(d)$ principal bundle $\widetilde{E}_\rho$ such that the following commutes up to homotopy
\begin{equation}
    \begin{tikzcd}
        & BSpin(d) \arrow[d, "B\pi"] \\
        BG \arrow[r, "f"'] \arrow[ru, "\tilde{f}", dashed] & BSO(d)
    \end{tikzcd}
\end{equation}
where $f:BG\rightarrow BSO(d)$ and $\widetilde{f}:BG\rightarrow BSpin(d)$ are the maps defining the bundles  $E_\rho$ and $\widetilde{E}_\rho$ by pullback, and $B\pi$ is the natural projection induced by the double cover $\pi :Spin(d)\rightarrow SO(d)$.

The holonomy of the lifted bundle $\widetilde{E}_\rho$ is by construction a representation $\tilde{\rho}:G\rightarrow Spin(d)$ such that
\begin{equation}\label{eq:commtriangle}
    \pi \circ \tilde{\rho} = \rho\,.
\end{equation}
Then in the diagram \eqref{eq:pullback} one can add $\tilde{\rho}$ as follows:
\begin{equation}
    \begin{tikzcd}
        1 \arrow[r] & \mathbb{Z}_2 \arrow[r] \arrow[d, equal] & G_F \arrow[r,"\pi_G"] \arrow[d, "\rho_F"'] & G \arrow[r] \arrow[d, "\rho"] \arrow[ld, "\tilde{\rho}"] & 1 \\
        1 \arrow[r] & \mathbb{Z}_2 \arrow[r] & Spin(d) \arrow[r, "\pi"'] & SO(d) \arrow[r] & 1
    \end{tikzcd}\,.
\end{equation}
Where only the bottom triangle commutes, not the top one.

Finally, we can use this diagram to build a section $s:G\mapsto G_F$, thus proving that the extension is split.

Consider an element $g\in G$. Its preimage $\pi_G^{-1}(g)$ in $G_F$ consists of two elements $f_1,f_2$ by exactness of the upper sequence. $\rho_F$ is faithful, so the images $\rho_F(f_1)$ and $\rho_F(f_2)$ are different, but they both project into the same $SO(d)$ element $\rho(g)$ by commutativity of the second square. By commutativity \eqref{eq:commtriangle}, exactly one of $\rho_F(f_1)$ and $\rho_F(f_2)$ will coincide with $\tilde{\rho}(g)$, say WLOG that it is $f_1$. Then define the section $s$ through $s(g)=f_1$. One can straightforwardly show that this is a group homomorphism $G\rightarrow G_F$, the unique one such that
\begin{equation}
    s\circ \rho_F = \tilde{\rho}\,.
\end{equation}
Since we have built a section $s:G\rightarrow G_F$, we conclude that the extension splits and $G_F \simeq G_B \times \Z_2$.

Conversely, if we were to deal with a non-trivial extension $G_F \neq G \times \Z_2$, we can immediately conclude that $w_2(\rho)\neq 0$.

\section{Subdirect products}\label{app:subdirect}

In this Appendix we explain how subdirect products arise in our fermionic group search, and how to construct them systematically.

When looking for the full orbifold group, what happens operationally is the following. One starts from the bosonic groups $G_B^L$ and $G_B^R$, which are crystallographic, i.e.~finite subgroups of $GL(d;\Z)$. The corresponding fermionic actions on the left and on the right are, separately, subgroups of $SU(4)$, isomorphic as finite groups to $G^{L/R}_B\times \Z_2$. We build the full fermionic action on the 8-dimensional ``$L+R$" space by taking direct sums of the generators of $G_F$.

Say we take two sets of generators $\{g_i^L\}_{i=1}^N$ and $\{g_i^R\}_{i=1}^N$ of the same size. Then we are building the orbifold group $G$ as the image of the map:
\begin{equation}
    (g_i^L,g_i^R) \longmapsto g_i^L \oplus g_i^R\,.
\end{equation}
This gives $G$ a natural embedding as a subgroup of the direct product $G^L\times G^R$. Furthermore, it is equipped with two surjective projections
\begin{equation}
    \begin{aligned}
        \pi_L: g_i^L \oplus g_i^R &\longmapsto g_i^L\,, \\
        \pi_R: g_i^L \oplus g_i^R &\longmapsto g_i^R\,.
    \end{aligned}
\end{equation}
These properties define what is called a subdirect product of $G_F^L$ and $G_F^R$, which we denote $\widetilde{G}=G_F^L \oplus G_F^R$.

Goursat's lemma \cite{ASENS_1889_3_6__9_0} gives a way to construct all possible subdirect products $\widetilde{G}$ of two given groups $G_1,G_2$ as fiber products, which is more amenable to an algorithmic implementation.

The statement is that there is a bijection between the set of subdirect products of two groups $G_1$ and $G_2$ and triplets
\begin{equation}
    (N_1,N_2,\phi)\,,
\end{equation}
where $N_1$ is a normal subgroup of $G_1$, $N_2$ is a normal subgroup of $G_2$, and $\phi:G_1/N_1 \rightarrow G_2/N_2$ is an isomorphism between the induced quotient groups.

The lemma gives an operative way to construct these subdirect products: one lists all normal subgroups of $G_F^L$ and $G_F^R$ and pairs them up in such a way that the resulting quotient groups $G_i/N_i$ are isomorphic. Then for every isomorphism $\phi$, one can build a subdirect product $G$ as a fiber product
\begin{equation}
    G \simeq G_1 \times_\phi G_2 := \left\{(g_1,g_2) \in G_1 \times G_2 \,\big{|}\,\phi(q_1(g_1)) = q_2(g_2) \right\}\,,
\end{equation}
where $q_i: G_i\rightarrow G_i/N_i$ is the canonical quotient map.

\bibliographystyle{JHEP.bst}
\bibliography{bibliography}

\end{document}